\documentclass[12pt, draftclsnofoot, onecolumn,comsoc]{IEEEtran}

\usepackage{amsmath,graphicx}
\usepackage{amssymb}
\usepackage{subfigure}
\usepackage{graphicx}
\usepackage{graphics}
\usepackage{epstopdf}
\usepackage{amsmath}
\usepackage{multirow}
\usepackage{bm}
\usepackage{enumerate}
\usepackage{mathdots}
\usepackage{todonotes}
\usepackage{algorithmicx}
\usepackage{algpseudocode}
\usepackage{algorithm}
\usepackage{verbatim}
\usepackage{balance}
\usepackage{cite}
\usepackage{url}
\usepackage{booktabs}
\usepackage{tabularx}

\usepackage{color}
\definecolor{purple(x11)}{rgb}{0.63, 0.36, 0.94}
\definecolor{cadmiumgreen}{rgb}{0.0, 0.42, 0.24}

\newcommand{\vect}{\mathop{\mathrm{vec}}}

\newcommand{\supp}{\mathop{\mathrm{supp}}}
\newcommand{\blkdiag}{\mathop{\mathrm{blkdiag}}}
\newcommand{\trace}{\mathop{\mathrm{trace}}}

\usepackage{pifont}

\newcommand{\Nr}{N_{\mathrm{r}}}
\newcommand{\Nt}{N_{\mathrm{t}}}
\newcommand{\Ns}{N_{\mathrm{s}}}

\newcommand{\Gr}{G_{\mathrm{r}}}
\newcommand{\Gt}{G_{\mathrm{t}}}
\newcommand{\Lt}{L_{\mathrm{t}}}

\newcommand{\fmtr}{{\mathbf{f}}_{\text{tr}}^{(m)}}
\newcommand{\Fmtr}{{\mathbf{F}}_{\text{tr}}^{(m)}}

\newcommand{\wmtr}{{\mathbf{w}}_{\text{tr}}^{(m)}}
\newcommand{\Wmtr}{{\mathbf{W}}_{\text{tr}}^{(m)}}
\newcommand{\Wmtrc}{{\mathbf{W}}_{\text{tr}}^{(m)*}}

\newcommand{\Nc}{N_{\mathrm{c}}}
\newcommand{\prc}{p_{\mathrm{rc}}}
\newcommand{\ar}{{\mathbf{a}}_{\mathrm{R}}}
\newcommand{\at}{{\mathbf{a}}_{\mathrm{T}}}

\newcommand{\AR}{{\mathbf{A}}_{\mathrm{R}}}
\newcommand{\AT}{{\mathbf{A}}_{\mathrm{T}}}

\newcommand{\jj}{{\mathrm{j}}}
\newcommand{\cH}{\mathbf{H}}
\newcommand{\FRF}{{\mathbf{F}}_{\mathrm{RF}}}
\newcommand{\FBB}{{\mathbf{F}}_{\mathrm{BB}}}
\newcommand{\WRF}{{\mathbf{W}}_{\mathrm{RF}}}
\newcommand{\WBB}{{\mathbf{W}}_{\mathrm{BB}}}

\newcommand{\be}{\begin{eqnarray}}
\newcommand{\ee}{\end{eqnarray}}

\def\ba{{\mathbf{a}}}
\def\bb{{\mathbf{b}}}
\def\bc{{\mathbf{c}}}

\def\bh{{\mathbf{h}}}

\def\bn{{\mathbf{n}}}

\def\bp{{\mathbf{p}}}

\def\br{{\mathbf{r}}}
\def\bs{{\mathbf{s}}}
\def\bt{{\mathbf{t}}}

\def\bx{{\mathbf{x}}}

\def\bz{{\mathbf{z}}}
\def\b0{{\mathbf{0}}}

\def\bA{{\mathbf{A}}}
\def\bB{{\mathbf{B}}}
\def\bC{{\mathbf{C}}}
\def\bD{{\mathbf{D}}}

\def\bH{{\mathbf{H}}}
\def\bI{{\mathbf{I}}}

\def\bW{{\mathbf{W}}}
\def\bX{{\mathbf{X}}}

\def\sf0{{\mathsf{0}}}

\def\bsfn{{\bm{\mathsf{n}}}}

\def\bsfr{{\bm{\mathsf{r}}}}
\def\bsfs{{\bm{\mathsf{s}}}}

\def\bsfx{{\bm{\mathsf{x}}}}
\def\bsfy{{\bm{\mathsf{y}}}}

\def\bsf0{{\bm{\mathsf{0}}}}

\begin{document}
\title{Frequency-domain Compressive Channel Estimation for Frequency-selective  Hybrid mmWave MIMO Systems}
\author{Javier Rodr\'{i}guez-Fern\'{a}ndez$^{\dag}$, Nuria Gonz\'{a}lez-Prelcic$^{\dag}$,  Kiran Venugopal$^{\ddag}$, and Robert W. Heath Jr.$^{\ddag}$ \thanks{This work was partially funded by the Agencia Estatal de Investigaci�n (Spain) and the European Regional Development Fund (ERDF) under project  MYRADA (TEC2016-75103-C2-2-R),  the U.S. Department of Transportation through the Data-Supported Transportation Operations and Planning (D-STOP) Tier 1 University Transportation
Center, by the Texas Department of Transportation under Project 0-6877 entitled Communications and Radar-Supported Transportation Operations and Planning (CAR-STOP) and by the National Science Foundation under Grant NSF-CCF-1319556 and NSF-CCF-1527079.}
\\
$^\dag$ Universidade de Vigo, Email: $\{$jrodriguez,nuria$\}$@gts.uvigo.es \\
$^\ddag$ The University of Texas at Austin, Email: $\{$kiranv,rheath$\}$@utexas.edu}

\maketitle

\begin{abstract}
Channel estimation is useful in millimeter wave (mmWave) MIMO communication systems. Channel state information 
allows optimized designs of precoders and combiners under different metrics such  as mutual information or signal-to-interference-noise (SINR) ratio.
At mmWave, MIMO precoders and combiners are usually hybrid,  since this architecture provides a means to trade-off power consumption and achievable rate. Channel estimation is challenging when using these architectures, however, since there is no direct access to the outputs of the different antenna elements in the array.
The MIMO channel can only be observed through the analog combining network, which acts as a  compression stage of the received signal.
Most of prior work on channel estimation for hybrid architectures assumes a frequency-flat mmWave channel model. In this paper, we consider a frequency-selective mmWave channel and propose compressed-sensing-based strategies to estimate the channel in the frequency domain.
We evaluate different algorithms and compute their complexity to expose trade-offs in complexity-overhead-performance as compared to those of previous approaches.
 \end{abstract}


\section{Introduction}

MIMO architectures with large arrays are a key ingredient of  mmWave communication systems, providing gigabit-per-second data rates \cite{mmWaveBook}.
Hybrid MIMO structures have been proposed to operate at mmWave because the cost and power consumption of an all-digital achitecture is prohibitive at these frequencies \cite{HeaPreRanRohSay:Overview-Sig-Proc-mmWaveMIMO:16}. Optimally configuring the digital and analog precoders and combiners requires channel knowledge when the design goal is maximizing metrics such as the achievable rate or the SINR. Acquiring the mmWave channel is challenging with a hybrid architecture, however, because the channel is seen through the analog combining network, the SNR is low before beamforming, and the size of the channel matrices is large \cite{Ahmed_MIMOprecoding_combining:CM2014}.


A significant  number of papers have  proposed solutions to the problem of  channel estimation at mmWave with hybrid architectures, but under a frequency-flat channel model  \cite{Malloy2012,Malloy2012a,Iwen2012, AlkAyaLeuHea:Channel-Estimation-and-Hybrid:14,Ramasamy2012a,Ramasamy2012b,Berraki2014,Lee2014,RiaRusPreAlkHea:Hybrid-MIMO-Architectures:16}. These strategies exploit
the spatially sparse structure in the mmWave MIMO channel, formulating the estimation of the channel as a sparse recovery problem. The support of the 
estimated sparse vector identifies the pairs of direction-of-arrival/direction-of-departure (DoA/DoD) for each path in the mmWave channel, while the amplitudes of the non-zero coefficients provide the channel gains for each path. Compressive estimations lead to a reduction in the channel training length when compared to conventional approaches such as those based on least squares (LS) estimation \cite{RiaRusPreAlkHea:Hybrid-MIMO-Architectures:16}.


Recently, some approaches for channel estimation in frequency-selective mmWave channels have been proposed.
In a recent paper \cite{VenAlkPreHeath:Time-domain-chan-estimation-wideband-hybrid:16}, we designed a time-domain approach to estimate the wideband mmWave channel assuming a hybrid MIMO architecture. This algorithm exploits the sparsity of the wideband millimeter wave channel in both the angular and delay domains. The sparse formulation of the problem in \cite{VenAlkPreHeath:Time-domain-chan-estimation-wideband-hybrid:16} includes the effect of non-integer sampling of the transmit pulse shaping filter, with the subsequent leakage effect and increase  of sparsity level in the channel matrix. The main limitation of \cite{VenAlkPreHeath:Time-domain-chan-estimation-wideband-hybrid:16} is that the algorithm can be applied only to single carrier systems. A frequency-domain strategy to estimate frequency-selective mmWave channels was also proposed in \cite{VenAlkGon:Channel-Estimation-for-Hybrid:17}. A sparse reconstruction  problem was formulated there to estimate the channel independently for every subcarrier, without exploiting spatial congruence between subbands. Another approach in the frequency domain was designed in \cite{RodGonVen:A-Frequency-Domain-Approach-to-Wideband:17}, but only exploting the information from a reduced number of subcarriers.
A different algorithm operating in the frequency domain for a MIMO-OFDM system was proposed in 
\cite{GaoDaiWan:Channel-estimation-mmWave-massiveMIMO:16}. Exploiting the fact that spatial propagation characteristics do not change significantly within the system bandwidth, \cite{GaoDaiWan:Channel-estimation-mmWave-massiveMIMO:16} assumed spatially common sparsity between the channels corresponding to the different subcarriers. A structured sparse recovery algorithm was then considered to reconstruct the channels in the frequency domain. Thus, \cite{GaoDaiWan:Channel-estimation-mmWave-massiveMIMO:16} is an interesting initial solution to the problem, but has several limitations when applied to a mmWave communication system: 

\begin{enumerate}
\item The effect of sampling the pulse shaping filter delayed by a non integer factor was not considered in the channel model for a given delay tap. As shown in \cite{schniter_sparseway:2014},  not accounting for this effect leads to virtual MIMO matrices with an artificially enhanced sparsity. 
\item The algorithm was evaluated only for medium and high SNR regimes (larger than 10 dBs), very unlikely at mmWave, where the expected SNR is below 0 dB. 
\item The reconstruction algorithm provides accurate results when Gaussian measurement matrices are employed; to generate the Gaussian matrices, unquantized phases were considered in the training precoders, which is unrealistic in a practical implementation of a mmWave system based on a hybrid architecture.
\item The algorithm is based on the strong assumption that the SNR is perfectly known at the receiver before explicit channel estimation. 
\end{enumerate}

Another algorithm exploiting common sparsity in the frequency domain at mmWave was proposed in \cite{GaoHuDai:Channel-estimation-mmWave-massiveMIMO-FS:16}. Unlike the algorithm proposed in \cite{GaoDaiWan:Channel-estimation-mmWave-massiveMIMO:16}, the algorithm in \cite{GaoHuDai:Channel-estimation-mmWave-massiveMIMO-FS:16} is proposed to estimate mmWave wideband MU-MIMO channels. Besides the limitations 1)-3) described above, which also hold in this case, this algorithm exhibits another problem that makes it less feasible to be applied in a real mmWave communication system. A Line-of-Sight (LoS) Rician channel model with $K_\text{factor} = 20$ dB is considered, which is only applicable when there is a strong LoS path. Owing to this artifact of the channel model, the algorithm in \cite{GaoHuDai:Channel-estimation-mmWave-massiveMIMO-FS:16} estimates a single path for each user, such that the task of channel estimation can not be successfully accomplished.

This paper proposes two novel frequency-domain approaches for the estimation of frequency-selective mmWave MIMO channels. These approaches overcome the limitations of prior work and provide different trade-offs between complexity and achievable rate for a fixed training length. As in recent work on hybrid architectures for frequency-selective mmWave channels \cite{GaoDaiWan:Channel-estimation-mmWave-massiveMIMO:16,AlkHea:Frequency-Selective-Hybrid:16}, we also consider a MIMO-OFDM communications system. Similar to \cite{VenAlkGon:Channel-Estimation-for-Hybrid:17}, we introduce zero-padding (ZP) as a means to avoid loss and/or distortion of training data during reconfiguration of RF circuitry.  A geometric channel is considered to model the different scattering clusters as in \cite{VenAlkPreHeath:Time-domain-chan-estimation-wideband-hybrid:16},\cite{GaoHuDai:Channel-estimation-mmWave-massiveMIMO-FS:16},\cite{schniter_sparseway:2014}, including the band-limiting property in the overall channel response. 

Our two proposed approaches exploit the spatially common sparsity within the system bandwidth. The first algorithm aims at exploiting the information on the support coming from every subcarrier in the MIMO-OFDM system and provides the best performance. In contrast, the second algorithm uses less information to estimate the different frequency-domain subchannels, thereby managing to significantly reduce the computational complexity. We show that both strategies are asymptotically efficient, since they both attain the Cram\'{e}r-Rao Lower Bound (CRLB). Further, we show that asymptotic efficiency can be achieved without using frequency-selective baseband precoders and combiners during the training stage, thereby reducing computational complexity. 

Simulation results in the low SNR regime show that the two proposed algorithms significantly outperform the approach in the frequency domain developed in \cite{VenAlkGon:Channel-Estimation-for-Hybrid:17}. Comparisons with the algorithms proposed in\cite{GaoDaiWan:Channel-estimation-mmWave-massiveMIMO:16} and \cite{GaoHuDai:Channel-estimation-mmWave-massiveMIMO-FS:16} are also provided to show their performance in terms of estimation error in the SNR regime that mmWave systems are expected to work. The two proposed channel estimation algorithms provide a good trade-off between performance and overhead. Results show that using a reasonably small  training length, approximately in the range of $60-100$ frames, leads to low estimation errors. The computational complexity of the proposed algorithms and previous strategies is also analyzed to compare the trade-offs between performance-complexity provided by the different algorithms. Finally, we also show that it is not necessary to exploit the information on the support coming from every OFDM subcarrier to estimate the different mmWave subchannels. Yet, a reduced number of subcarriers is enough to asymptotically attain the CRLB.

The paper is organized as follows: Section~\ref{section:systemmodel} introduces the system model that is used throughout this work. 
Section~\ref{section:compressivechannelestimation} proposes two frequency-domain compressive channel estimation approaches, including the computation of the CRLB and suitable estimators derived from it in order to obtain the different frequency-domain channel matrices for each subcarrier. Thereafter, Section~\ref{section:results} provides the main simulation results for the two proposed algorithms, the OMP-based compressive approach proposed in \cite{VenAlkPreHeath:Time-domain-chan-estimation-wideband-hybrid:16} and the SSAMP and DGMP algorithms proposed in \cite{GaoDaiWan:Channel-estimation-mmWave-massiveMIMO:16} and \cite{GaoHuDai:Channel-estimation-mmWave-massiveMIMO-FS:16}, respectively. Finally, Section~\ref{section:conclusions} collects the main conclusions derived from the simulation results and also describes future work to be conducted.

\textbf{Notation}: We use the following notation throughout this paper: bold uppercase $\bA$ is used to denote matrices, bold lowercase $\ba$ denotes a column vector and non-bold lowercase $a$ denotes a scalar value. We use ${\cal A}$ to denote a set. Further, $\|\bA\|_F$ is the Frobenius norm and $\bA^*$, $\overline{\bA}$, $\bA^T$ and $\bA^\dag$ denote the conjugate transpose, conjugate, transpose and Moore-Penrose pseudoinverse of a matrix $\bA$, respectively. The $(i,j)$-th entry of a matrix $\bA$ is denoted using $\{\bA\}_{i,j}$. Similarly, the $i$-th entry of a column vector $\ba$ is denoted as $\{\ba\}_i$. The identity matrix of order $N$ is denoted as $\bI_N$. If $\bA$ and $\bB$ are two matrices, $\bA \circ \bB$ is the Khatri-Rao product of $\bA$ and $\bB$ and $\bA \otimes \bB$ is their Kronecker product. We use ${\cal N}(\mathbf{\mu},\bC)$ to denote a circularly symmetric complex Gaussian random vector with mean $\mathbf{\mu}$ and covariance matrix $\bC$. We use $\mathbb{E}$ to denote expectation. Discrete-time signals are represented as $\bx[n]$, while frequency-domain signals are denoted using $\bsfx[k]$.

\section{System model}
\label{section:systemmodel}
\begin{figure*}[t!]
\centering
\includegraphics[width=0.8\textwidth]{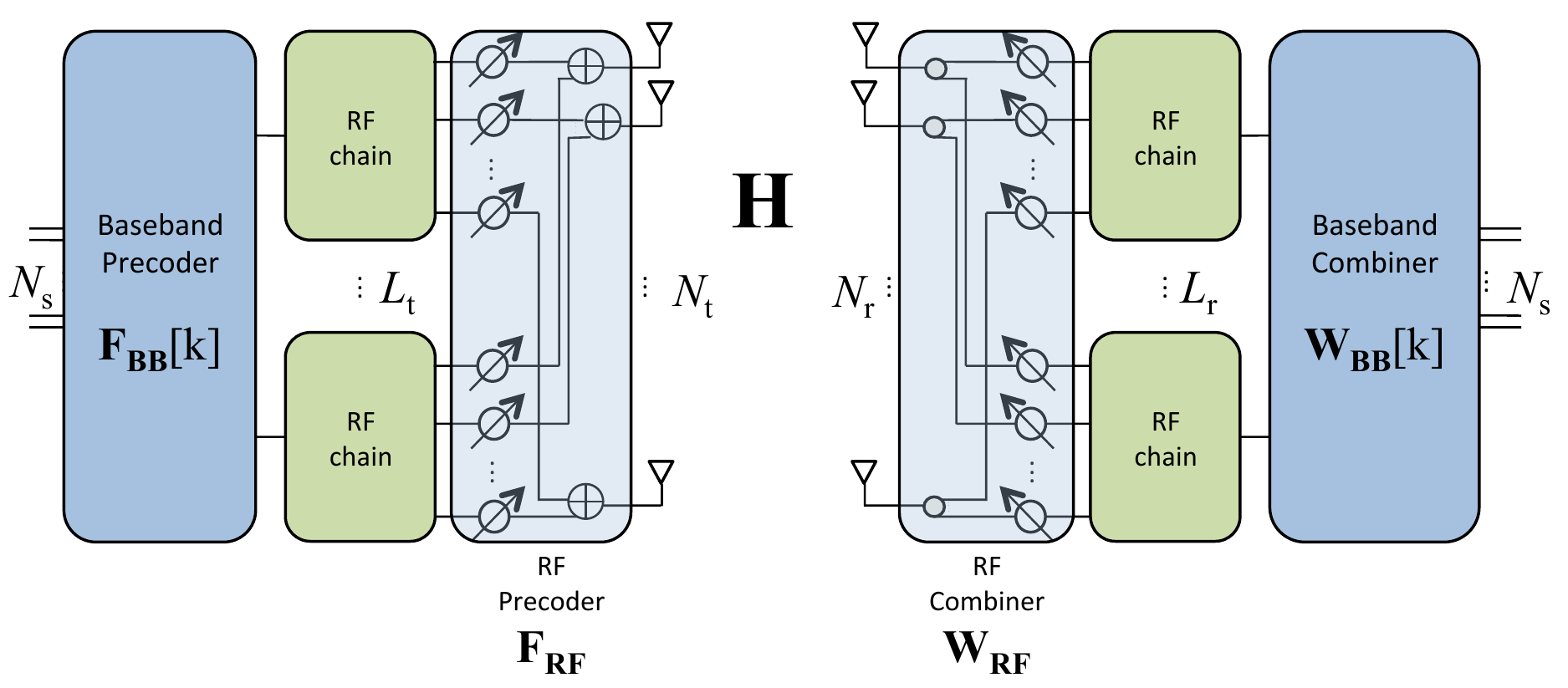} 
\caption{Illustration of the structure of a hybrid MIMO architecture, which include analog and digital precoders and combiners.}     
\label{fig:hybrid_architecture}        
\end{figure*}

We consider an OFDM based mmWave MIMO link employing $K$ subcarriers to send $\Ns$ data streams using a transmitter with $\Nt$ antennas and a receiver with $\Nr$ antennas. The system is based on a hybrid MIMO architecture as shown in Fig. \ref{fig:hybrid_architecture}, with $L_\text{t}$ and $L_\text{r}$ RF chains at the transmitter and receiver sides. For a general exposition, a frequency-selective hybrid precoder is used, with $\mathbf{F} [k]= \FRF\FBB[k] \in {\mathbb{C}}^{\Nt\times\Ns}$, $k=0,\ldots,K-1$, where $\FRF$ is the analog precoder and $\FBB[k]$ the digital one. Note that the analog precoder is frequency-flat, while the digital precoder is different for every subcarrier. The RF precoder and combiner are implemented using a fully connected network of phase shifters, as described in \cite{RiaRusPreAlkHea:Hybrid-MIMO-Architectures:16}. The symbol blocks are transformed into the time domain using $L_\text{r}$ parallel $K$-point IFFTs. 
As in \cite{VenAlkGon:Channel-Estimation-for-Hybrid:17,LarThoCud:Air-interface-design-and-ray-tracing:13}, we consider Zero-Padding (ZP) to both suppress Inter Symbol Interference (ISI) and account for the RF circuitry reconfiguration time. The discrete-time complex baseband signal at subcarrier $k$ can be written as
\begin{equation}
\bsfx[k] = \FRF\FBB[k] \bsfs[k], 
\end{equation}
where the transmitted symbol sequence at subcarrier $k$ of size $\Ns \times 1$ is denoted as $\bsfs[k]$.

The MIMO channel between the transmitter and the receiver is assumed to be frequency-selective, having a delay tap length $\Nc$ in the time domain.
The $d$-th delay tap of the channel is represented
by a $\Nr \times \Nt$ matrix denoted as $\cH_d,~d=0,~1,~...,~\Nc-1$, which, assuming a geometric channel model \cite{schniter_sparseway:2014}, can be written as
\be
\hspace*{-3.5mm}\cH_d &=\hspace*{-1mm}& \sqrt{\frac{N_\text{t} N_\text{r}}{L\rho_\text{L}}}\sum_{\ell = 1}^{L}\alpha_{\ell}\prc(dT_\text{s}-\tau_{\ell})\ar(\phi_{\ell})\at^*(\theta_{\ell}), \label{eqn:channel_model}
\ee
where $\rho_\text{L}$ denotes the path loss between the transmitter and the receiver, $L$ denotes the number of paths, $\prc(\tau)$ is a filter that includes the effects of pulse-shaping and other lowpass filtering evaluated at $\tau$, $\alpha_{\ell} \in {\mathbb{C}}$ is the complex gain of the $\ell$th path, $\tau_{\ell} \in {\mathbb{R}}$ is the delay of the $\ell$th path, $\phi_{\ell} \in [0, 2\pi)$ and $\theta_{\ell} \in [0, 2\pi)$ are the angles-of-arrival and departure (AoA/AoD), of the $\ell$th path, and $\ar(\phi_{\ell}) \in {\mathbb{C}}^{\Nr\times1}$ and $\at(\theta_{\ell}) \in {\mathbb{C}}^{\Nt\times1}$ are the array steering vectors for the receive and transmit antennas.  
Each one of these matrices can be written in a more compact way as
\be
\cH_d &=& \AR \mathbf{\Delta}_d\AT^*, \label{eqn:channel_compact}
\ee where $\mathbf{\Delta}_d \in {\mathbb{C}}^{L\times L}$ is diagonal with non-zero complex entries, 
and $\AR \in {\mathbb{C}}^{\Nr\times L}$ and $\AT \in {\mathbb{C}}^{\Nt\times L}$ contain the receive and transmit array steering vectors $\ar(\phi_{\ell})$ and $\at(\theta_{\ell})$, respectively. The channel $\cH_d$ can be approximated using the extended virtual channel model defined in  \cite{HeaPreRanRohSay:Overview-Sig-Proc-mmWaveMIMO:16} as
\begin{equation}
			\cH_d \approx
				 \tilde{\bA}_\text{R} \mathbf{\Delta}_d^v \tilde{\bA}_\text{T}^*,
\label{eq:virtual_ch_model}
\end{equation}
where  $\mathbf{\Delta}_d^v \in \mathbb{C}^{\Gr \times \Gt}$ is a sparse matrix which contains the path gains of the quantized spatial frequencies in the non-zero elements. The dictionary matrices  $\tilde{\bA}_\text{T}$ and $\tilde{\bA}_\text{R}$  contain the transmitter and receiver array response vectors evaluated on a grid of size $\Gr$ for the AoA and a grid of size $\Gt$ 
for the AoD. Due to the few scattering clusters in mmWave channels, the sparse assumption for $\mathbf{\Delta}_d^v \in \mathbb{C}^{\Gr \times \Gt}$ is commonly accepted.

Finally, the channel at subcarrier $k$ can be written in terms of the different delay taps as
\be
\bH[k]=\sum_{d=0}^{\Nc-1} \cH_d e^{-\jj\frac{2\pi k}{K}d}.
\ee
It is also useful to write this matrix in terms of the sparse matrices $\mathbf{\Delta}_d^v$
and the dictionaries
\be
\bH[k] \approx \tilde{\bA}_\text{R}  {\bigg{(}}\sum_{d=0}^{\Nc-1}  \mathbf{\Delta}_d^v  e^{-\jj\frac{2\pi k}{K}d} {\bigg{)}} \tilde{\bA}_\text{T}^*
=  \tilde{\bA}_\text{R} \mathbf{\Delta}[k] \tilde{\bA}_\text{T}^*
\label{equation:final_H}
\ee
to help expose the sparse structure later.

Assuming that the receiver applies a hybrid combiner ${\mathbf{W}}[k]={\WRF\WBB[k]} \in {\mathbb{C}}^{\Nr\times N_\text{s}}$, the received signal at subcarrier $k$ can be written as
\be
\begin{split}
\bsfy[k]  =  \WBB^*[k]\WRF^*\bH[k]\FRF\FBB[k]\bsfs[k]\\
 + \WBB^*[k]\WRF^*\bsfn[k],\\
\end{split}
\label{equation:signal_model}
\ee
where $\bsfn[k] \sim \mathcal{N}\left(0,\sigma^2 \mathbf{I}\right)$  is the circularly symmetric complex Gaussian distributed additive noise vector. The receive signal model in \eqref{equation:signal_model} corresponds to the data transmission phase. As we will see in Section \ref{section:compressivechannelestimation}, during the 
channel acquisition phase, we will consider analog-only training precoders and combiners to reduce complexity. 
\section{Compressive Channel Estimation in the Frequency Domain}
\label{section:compressivechannelestimation}

In this section, we formulate a compressed sensing problem to estimate the vectorized
sparse channel vector in the frequency domain. We also propose two algorithms to solve this problem that leverage the common support between the channel matrices for every subcarrier, providing different trade-offs complexity-performance. The first algorithm leverages the common support  between the $K$ different subchannels providing a very good performance, while the second one only exploits information from a reduced number of subcarriers, thereby keeping computational complexity at a lower level.

\subsection{Problem formulation}
We assume that $L_\text{t}$ RF chains are used at the transmitter. During the training phase, for the $m$-th frame we use a training precoder $\Fmtr$ and a training combining matrix $\Wmtr$. This means that during the training phase, only analog precoders and combiners are considered to keep the complexity of the sparse recovery algorithm low. We assume that the transmitted symbols satisfy $\mathbb{E}\{\bsfs^{(m)}[k]\bsfs^{(m)*}[k]\} = \frac{P}{N_\text{s}}\bI_{N_\text{s}}$, with $P$ the total transmitted power and $\Ns = \Lt$. Furthermore, each entry in $\Fmtr$, $\Wmtr$ is normalized to have squared-modulus $\Nt^{-1}$ and $\Nr^{-1}$, respectively. Then, the received samples in the frequency domain for the $m$-th training frame can be written as
\begin{equation}
	\bsfy^{(m)}[k] = {\Wmtr}^* \bH[k] \Fmtr \bsfs^{(m)}[k] + \bsfn_\text{c}^{(m)}[k],
	\label{eq:received_freq_domain}
\end{equation}
where $\bH[k] \in \mathbb{C}^{N_r\times N_t}$ is the frequency-domain MIMO channel response at the $k$-th subcarrier and $\bsfn_\text{c}^{(m)}[k] \in \mathbb{C}^{L_r\times 1}$, $\bsfn_\text{c}^{(m)}[k] = {\Wmtr}^*\bsfn^{(m)}[k]$, is the frequency-domain combined noise vector received at the $k$-th subcarrier. The average received SNR per transmitted symbol is given by $\text{SNR} = \frac{P}{\rho_\ell \sigma^2}$. We assume that the channel coherence time is larger than the frame duration and that the same channel can be considered for several consecutive frames. 

Using the result $\vect\{\bA\bX\bC\} = (\bC^T \otimes \bA)\vect\{\bX\}$, the vectorized received signal  is
\begin{equation}
	\vect\{\bsfy^{(m)}[k]\} = (\bsfs^{(m)T}[k]{\Fmtr}^T \otimes {\Wmtr}^*)\vect\{\bH[k]\} + \bsfn_\text{c}^{(m)}[k].
	\label{eq:vectorized_freq_domain}
\end{equation}
Taking into account the expression in \eqref{equation:final_H}, the vectorized channel matrix can be written as $\vect\{\bH[k]\} = (\bar{\tilde{\bA}}_\text{T} \otimes \tilde{\bA}_\text{R} )\vect\{\mathbf{\Delta}[k]\}$. Therefore, if we define the measurement matrix $\mathbf{\Phi}^{(m)}[k] \in \mathbb{C}^{L_\text{r} \times N_\text{t} N_\text{r}}$ as
\begin{equation}
	\mathbf{\Phi}^{(m)}[k] = (\bsfs^{(m)T}[k]{\Fmtr}^T\otimes {\Wmtr}^*)
\end{equation} 
and the dictionary $\mathbf{\Psi} \in \mathbb{C}^{N_\text{t} N_\text{r} \times G_\text{t} G_\text{r}}$ as
\begin{equation}
	\mathbf{\Psi} = (\bar{\tilde{\bA}}_\text{T} \otimes \tilde{\bA}_\text{R}),
\end{equation}
(\ref{eq:vectorized_freq_domain}) can be rewritten as
\begin{equation}
	\vect\{\bsfy^{(m)}[k]\} = \mathbf{\Phi}^{(m)}[k] \mathbf{\Psi}\bh^\text{v}[k] + \bsfn_\text{c}^{(m)}[k],
	\label{eq:received_signal_M}
\end{equation}
where $\bh^\text{v}[k] = \vect\{\mathbf{\Delta}[k]\} \in \mathbb{C}^{G_\text{r}G_\text{t}\times 1}$ is the sparse vector containing the complex channel gains. 
To have enough measurements and accurately reconstruct the sparse vector $\bh^\text{v}[k]$, it is necessary to use several training frames, especially in the very-low SNR regime. If the transmitter and receiver communicate during $M$ training steps using different pseudorandomly built precoders and combiners, (\ref{eq:received_signal_M}) can be extended to
\begin{equation}
	\underbrace{\left[\begin{array}{cccc}
	\bsfy^{(1)}[k] \\
	\bsfy^{(2)}[k] \\
	\vdots \\
	\bsfy^{(M)}[k] \\ \end{array}\right]}_{\bsfy[k]} = \underbrace{\left[\begin{array}{cccc} \mathbf{\Phi}^{(1)}[k] \\
	\mathbf{\Phi}^{(2)}[k] \\ 
	\vdots \\
	\mathbf{\Phi}^{(M)}[k] \\ \end{array}\right]}_{\bm \Phi[k]} \mathbf{\Psi} \bh^\text{v}[k] + \underbrace{\left[\begin{array}{c}
	\bsfn_\text{c}^{(1)}[k] \\
	\bsfn_\text{c}^{(2)}[k] \\
	\vdots \\
	\bsfn_\text{c}^{(M)}[k] \\ \end{array}\right]}_{\bsfn_\text{c}[k]}.
	\label{equation:received_signal_complete}
\end{equation}

Finally, the vector $\bh^\text{v}[k]$ can be found by solving the sparse reconstruction problem 
\begin{equation}
	\min \|\bh^\text{v}[k]\|_1 \qquad \text{subject to} \qquad \|\bsfy[k] - \mathbf{\Phi}[k]\mathbf{\Psi}\bh^\text{v}[k]\|_2^2 < \epsilon,
	\label{CS_problem}
\end{equation}
where $\epsilon$ is a tunable parameter defining the maximum error between the measurement and the received signal assuming the reconstructed channel between the transmitter and the receiver. Since the sparsity (number of channel paths) is usually unknown, this parameter can be set to the noise variance \cite{VenAlkPreHeath:Time-domain-chan-estimation-wideband-hybrid:16}. 

There are a great variety of algorithms to solve \eqref{CS_problem}. We could  use, for example,  Orthogonal Matching Pursuit (OMP) to find the sparsest approximation of the vectors containing the channel gains. Since the vector $\bh^\text{v}[k]$ depends on the frequency bin, it would be necessary to run the algorithm as many times as the number of subcarriers at which the MIMO channel response is to be estimated.
In the next subsections we consider an additional assumption to solve this problem, which avoids the need to run $K$ OMP algorithms in parallel as proposed in \cite{VenAlkGon:Channel-Estimation-for-Hybrid:17}. 

The formulation in \eqref{eq:received_freq_domain}-\eqref{CS_problem} is provided as a general framework to leverage the sparse structure of the frequency-selective mmWave channel. This formulation, however, leads to subcarrier-dependent measurement matrices, resulting in high computational complexity for channel estimation. Using more than a single RF chain at the transmitter offers more degrees of freedom to design the measurement matrices and, consequently, obtain better estimation performance. Nonetheless, exploiting these degrees of freedom results in additional computational complexity with regard to the case of using a single RF chain at the transmitter. Furthermore, the difference in performance between using $L_\text{t} = 1$ and $L_\text{t} > 1$ was observed to be less than $0.5$ dB when solving the problem with the algorithm proposed in this section, which comes from the fact that a reasonable number of RF chains must be used at a given transceiver to keep power consumption low. For this reason, our focus is on designing algorithms exhibiting reasonable performance and low computational complexity.




For this reason, we consider the special case of $\Lt = 1$ to present our algorithms with reduced complexity. With this choice, the frequency-selective transmitted symbols are scalar and its effect can be easily inverted at the receiver (i.e., by multiplying by $(s^{(m)}[k])^{-1}$ or multiplying by $\overline{s^{(m)}}[k]$ if the transmitted symbols belong to a QPSK constellation with energy-normalized symbols). As shown in Section~\ref{section:results}, the use of a single frequency-flat measurement matrix is enough to asymptotically attain the CRLB with on-grid channel parameters. Therefore, a single RF chain and analog-only precoders are used during training, thereby managing a reasonable trade-off between performance and computational complexity. Nonetheless, the algorithms we propose in the next subsections can be easily extended to work with subcarrier-dependent measurement matrices.



By compensating for $s^{(m)}[k]$, for a given training step $m$, the measurement matrix $\bm \Phi^{(m)}[k]$ reduces to $\bm \Phi^{(m)} = {\fmtr}^T \otimes \Wmtrc$. At this point, it is important to highlight that: 1) Analog-only training precoders and combiners are used to keep computational complexity low, and 2) A single data stream is transmitted to guarantee that the same measurement matrix is enough to estimate the different frequency-domain subchannels. In fact, these choices enable both online and offline complexity reduction when the noise statistics are used to estimate the MIMO channel at the different subcarriers.

The matrices $\mathbf{\Delta}[k]$ exhibit an interesting property that can be exploited when  solving the compressed channel estimation problems defined in \eqref{CS_problem}. 
Let us define the $G_\text{t}G_\text{r} \times 1$ vectorized virtual channel matrix for a given delay tap as
\begin{equation}
	\bh_d^\text{v} \triangleq \vect\{\mathbf{\Delta}_d^v\}.
\end{equation}
Let us denote the supports of the virtual channel matrices $\mathbf{\Delta}_d^v$ as ${\cal T}_0, {\cal T}_1, \ldots, {\cal T}_{N_\text{c}-1}$, $d = 0, \ldots, N_\text{c}-1$. Then, since $\bh^\text{v}[k] = \vect\{\mathbf{\Delta}[k]\}$, with $\mathbf{\Delta}[k] = \sum_{d=0}^{\Nc-1}  \mathbf{\Delta}_d^v  e^{-j\frac{2\pi k}{N}d}$, $k = 0,\ldots,K-1$, it is clear that
\begin{equation}
	\supp\{\bh^\text{v}[k]\} = \bigcup_{d=0}^{N_\text{c}-1}{\supp\{\bh_d^\text{v}\}} \qquad k = 0,\ldots,K-1,
\end{equation}
where the union of the supports of the time-domain virtual channel matrices comes from the additive nature of the Fourier transform. 
The sparse assumption on the vectorized  channel matrix for a given delay tap $\bh_d^\text{v}$ is commonly accepted, since in mmWave channels
$L<< G_\text{r}G_\text{t}$. The vectorized channel matrix $\bh^\text{v}[k]$ will have, in the worst case, $\Nc L$ non-zero coefficients.
Typical values for $N_\text{c}$ in mmWave channels are usually lower than $64$ symbols (for example IEEE 802.11ad has been designed to work robustly for a maximum of $64$ delay taps in the channel), while the number of measured paths usually satisfies $L<30$ 
for outdoor and indoor scenarios \cite{Rap:TCOM:2015}. From these values, using dictionaries of size $G_\text{r} \geq 64\; G_\text{t}\geq 64$, allows us to assume
a spatially sparse structure for $\bh^\text{v}[k]$ as well. Furthermore, this sparse structure is the same for all $k$, since from \eqref{equation:final_H}, $\tilde{\bA}_\text{R}$ and $\tilde{\bA}_\text{T}$ do not depend on $k$, which means that the AoAs and AoDs do not change with frequency in the transmission bandwidth.

\subsection{Simultaneous weighted estimation exploiting common support between subcarriers (SW-OMP)}

To develop a channel estimation algorithm that leverages the sparse nature and the common support property for all $\bh^\text{v}[k]$, we propose to modify the S-OMP algorithm proposed in \cite{TroGilStr:Simultaneous-sparse-approximation:05}. For a given iteration, this algorithm aims at finding a new index of the support such that its likelihood is larger than if the signals for the different subbands were individually processed. In this way, the $K$ different signals contribute to decide which is the most likely index belonging to the support in a cooperative fashion.
The S-OMP algorithm in \cite{TroGilStr:Simultaneous-sparse-approximation:05} computes the gains of the sparse vector using a LS approach once the support is obtained, assuming that the noise covariance matrix is the identity matrix $\bI_{ML_r}$. In this paper, we propose two modifications of the S-OMP for the estimation of the support and the channel gains. The first one accounts for the correlated nature of the noise at the output of the RF combiner, and leads to a more precise estimation of the support in this particular application. The second one consists of  a different approach to compute the channel gains. In particular, we develop the minimum variance unbiased estimator  for the channel gains, which is shown to be a weighted LS estimator. 
We show that this estimator attains the CRLB.

\subsubsection{Support computation with correlated noise}
Before explicit estimation of the channel gains, it is necessary to compute the atom, i.e., vector in the measurement matrix, which yields the largest sum-projection onto the received signals, provided that the support of the different sparse vectors is the same. The S-OMP algorithm is based on the assumption that the perturbation (noise) covariance matrix is diagonal, such that no correlation between the different noise components is present. The projection vector $\bc[k] \in \mathbb{C}^{M L_\text{r}}$ is defined as
\begin{equation}
    \bc[k] = \bm \Upsilon^* \bsfy[k],
    \label{projectionSOMP}
\end{equation}
in which $\bm \Upsilon \in \mathbb{C}^{M L_\text{r} \times G_\text{t} G_\text{r}}$, $\bm \Upsilon = \bm \Phi \bm \Psi$ is the measurement matrix and $\bsfy[k] \in \mathbb{C}^{M L_\text{r} \times 1}$ is the received signal for a given $k$, $k = 0,\ldots,K-1$. If there is correlation between noise components, the atom estimated from the projection in (\ref{projectionSOMP}) is not likely to be the actual one. To introduce the appropriate correction in the projection, the specific form the noise covariance matrix takes needs to be used. It is important to highlight that depending on the design of the combiners for each of the $M$ training steps, the noise statistics will be different. Specifically, the noise covariance matrix contains the (complex) inner product between the different hybrid combiners, for each training step $m$, $m = 1,\ldots,M$. Let us consider two arbitrary (hybrid) combiners ${\Wmtr}^{(i)}$, ${\Wmtr}^{(j)} \in \mathbb{C}^{N_\text{r} \times L_\text{r}}$ for two arbitrary training steps $i$,$j$ and a given subcarrier $k$. Therefore, if the combined noise at a given training step $i$ and subcarrier $k$ is denoted as $\bn_\text{c}^{(i)}[k] = {\Wmtr}^{(i)*}\bn^{(i)}[k]$, with $\bn^{(i)}[k] \sim {\cal N}(\bm 0,\sigma^2\bI_{L_\text{r}})$, the noise cross-covariance matrix is given by
\begin{align}
	\mathbb{E}\{\bsfn_\text{c}^{(i)}[k] \bsfn_\text{c}^{(j)*}[k]\} & = \mathbb{E}\{{\Wmtr}^{(i)*}\bsfn^{(i)}[k] \bsfn^{(j)*}[k] {\Wmtr}^{(j)}\} \nonumber \\
	& = {\Wmtr}^{(i)*} \mathbb{E}\{\bsfn^{(i)}[k] \bsfn^{(j)*}[k]\}{\Wmtr}^{(j)}\nonumber \\
	& = {\Wmtr}^{(i)*}\sigma^2\delta[i-j] {\Wmtr}^{(j)},
\end{align}
such that it is the same for all subcarriers during training. It can be written as a block diagonal matrix $\bC\in \mathbb{C}^{M L_\text{r} \times M L_\text{r}}$ whose $i$-th diagonal block is the Gram matrix of the (hybrid) combiner $\bW^{(i)}$, that is,  $\bC=\sigma^2 \bC_\text{w}$, where 
\begin{equation}
	\bC_\text{w}= \blkdiag\{{\Wmtr}^{(1)*} {\Wmtr}^{(1)},{\Wmtr}^{(2)*} {\Wmtr}^{(2)},\ldots,{\Wmtr}^{(M)*} {\Wmtr}^{(M)}\}.
\end{equation}

Now, we can resort to the Cholesky decomposition of $\bC_\text{w}$
\begin{equation}
    \bC_\text{w}= \bD_\text{w}^* \bD_\text{w},
\end{equation}
with $\bD_\text{w} \in \mathbb{C}^{M L_\text{r} \times M L_\text{r}}$ an upper triangular matrix. The subscript in $\bC_\text{w}$ and $\bD_\text{w}$ indicates that these matrices only depend on the combiners $\Wmtr$ used during $M$ consecutive training frames. Therefore, the projection step is performed as
\begin{equation}
    \bc[k] = \bm \Upsilon_\text{w}^* \bsfy[k],
\end{equation}
where $\bm \Upsilon_\text{w} \in \mathbb{C}^{M L_\text{r} \times G_\text{t} G_\text{r}}$ is the whitened measurement matrix given by $\bm \Upsilon_\text{w} = \bD_\text{w}^{-1}\bm \Upsilon$. In this way, the resulting projection simultaneously whitens the spatial noise components and estimates the projection of the received signal onto the subspace spanned by the columns in the measurement matrix.

\subsubsection{Computation of the channel gains}
\label{subsec:channel_gains}

If the vectorized channel matrix for a given subcarrier $k$, $k = 0,1,\ldots,K-1$, is given by $\vect\{\bH[k]\} = (\bar{\bA}_\text{T} \circ \bar{\bA}_\text{R})\bm \xi[k]$, with $\bm \xi[k]$ the $L\times 1$ vector of channel gains, then the received signal $\bsfy[k]$ is distributed according to ${\cal N}(\bm \Phi \bm \Psi \bm \xi[k],\bC)$. Once the different AoAs/AoDs are found, the signal model for the $k$-th subcarrier can be written as
\begin{equation}
	\bsfy[k] = \mathbf{\Upsilon}_{\{:,\hat{\cal T}\}} \tilde{\bm{\xi}}[k] + \tilde{\bsfn}_\text{c}[k],
\label{equation:linear_model}	
\end{equation}

where $\tilde{\bsfn}_\text{c}[k] \in \mathbb{C}^{M L_\text{r} \times 1}$ is the residual noise in our linear model after estimating the channel support. If the estimation of the support is accurate, $\tilde{\bsfn}_\text{c}[k]$ will be close to the post-combining noise vector $\bsfn_\text{c}[k]$. The $\hat{L} \times 1$ vector $\tilde{\bm \xi}[k]$ is the vector of channel gains to be estimated after sparse recovery, where $\hat{L} = |\hat{\cal T}|$ is the estimate of the sparsity level. The matrix $\mathbf{\Upsilon}_{\{:,\hat{\cal T}\}} \in \mathbb{C}^{M L_\text{r} \times \hat{L}}$ is defined as
\begin{equation}
	\mathbf{\Upsilon}_{\{:,\hat{\cal T}\}} = {\big{[}}\mathbf{\Phi} \mathbf{\Psi}{\big{]}}_{\{:,\hat{\cal T}\}}.
\end{equation}
 It is important to remark that the support estimated by the proposed algorithm may be different from the actual channel support. Provided that, in general, $\hat{\cal T}$ can be different from the actual support, the vector $\bm \xi[k] \in \mathbb{C}^{\hat{L}\times 1}$ containing the complex channel gains can be also different from the vector containing the true channel gains. 
Since the model in \eqref{equation:linear_model} is linear on the parameter vector $\bm \xi[k]$, there is a Minimum Variance Unbiased (MVU) Estimator that happens to be the Best Linear Unbiased Estimator (BLUE) as well \cite{Kay:Fundamentals-of-Statistical-Signal:93}.

The mathematical equation in (\ref{equation:linear_model}) is usually referred to as the \textit{General Linear Model} (GLM), for which the solution of $\tilde{\bm \xi}[k]$ for real parameters is provided in \cite{Kay:Fundamentals-of-Statistical-Signal:93}. The extension for a complex vector of parameters is straightforward and given by
\begin{equation}
	\hat{\tilde{\bm \xi}}[k] = (\mathbf{\Upsilon}_{\{:,\hat{\cal T}\}}^*\bC^{-1}\mathbf{\Upsilon}_{\{:,\hat{\cal T}\}})^{-1}\mathbf{\Upsilon}_{\{:,\hat{\cal T}\}}^*\bC^{-1}\bsfy[k].
\end{equation}
Therefore, $\hat{\tilde{\bm \xi}}[k]$ is the MVU estimator for our parameter vector $\tilde{\bm \xi}[k]$, $k = 0,\ldots,K-1$. Hence, it is unbiased and attains the CRLB is the support is estimated correctly. It is interesting to note that this corresponds to a Weighted Least-Squares (WLS) estimator, with the corresponding weights given by the inverse covariance matrix of the noise. We are interested in finding the CRLB for the estimation of the channel matrices at each subcarrier assuming perfect sparse reconstruction. To that end, taking into account only the non-zero entries in $\bh^\text{v}[k]$, the Fisher Information Matrix (FIM) is derived from the GLM in \eqref{equation:received_signal_complete} as
\begin{equation}
	\bI(\bm \xi[k]) = \mathbf{\Upsilon}_{\{:,{\cal T}\}}^*\bC^{-1} \mathbf{\Upsilon}_{\{:,{\cal T}\}}.
\end{equation}

Note that the previous expression gives the FIM for the vector $\bm \xi[k]$, which contains the actual channel gains.
Accordingly, the overall variance of the estimator for the vector of channel gains is given by the sum of the individual CRLBs for each of the complex gains. Therefore, if we denote the overall variance of the estimator for $\bm \xi[k]$ as $\gamma(\bm \xi[k])$, $k = 0,1,\ldots,K-1$, then 
\begin{equation}
	\text{CRLB}(\bm \xi[k]) = \bI^{-1}(\bm \xi[k])
\end{equation}
and
\begin{equation}
	\gamma(\bm \xi[k]) = \trace\{\text{CRLB}(\bm \xi[k])\}.
\end{equation}
The last equation only takes into account the channel gains for each path in the extended virtual channel representation. Our interest, however, is to compare the channel estimation performance at the antenna level. To that end, the frequency-domain channel matrix can be vectorized as (subcarrier-wise)
\begin{equation}
	\vect\{\bH[k]\} = (\overline{\bA_\text{T}} \circ \bA_\text{R}) \bm \xi[k],
	\label{equation:known_support_decomposition}
\end{equation}
where $\bA_\text{T} \in \mathbb{C}^{N_\text{t} \times L}$, $\bA_\text{R} \in \mathbb{C}^{N_\text{r} \times L}$ are the array response matrices evaluated on the AoAs/AoD. Notice that the decomposition in (\ref{equation:known_support_decomposition}) is expressed with equality, which follows from the assumption that the estimation of the support is perfect and only the channel gains are to be estimated. The assumption on the perfect estimation of the support comes from the fact that the channel vectors $\bh^\text{v}[k]$ are assumed to be sparse and the AoA/AoD are estimated perfectly. Since the Fisher Information requires that the mean of the received signal is continuous and differentiable on the parameters to estimate, it cannot be applied to the estimation of the angular parameters. Owing to our lacking space, the solution to the problem of finding the CRLB in a more general case accounting for continuous AoA and AoD is left for future work.

The overall minimum variance for an unbiased estimator of the $N_\text{t}N_\text{r}$ entries in $\bH[k]$ is given by the sum of the variances for each element $\{\bH[k]\}_{\{i,j\}}$, $i = 1,\ldots,N_\text{r}$, $j = 1,\ldots,N_\text{t}$. Mathematically, following the same notation as with $\gamma(\bm \xi[k])$, we will denote the overall variance of the estimator for $\bH[k]$ as $\gamma(\vect\{\bH[k]\})$, for a given subcarrier $k$. Then, $\gamma(\vect\{\bH[k]\})$ is derived as
\begin{eqnarray}
\gamma(\vect\{\bH[k]\})  &=& \trace{\bigg{\{}}\frac{\partial \vect\{\bH[k]\}}{\partial \bm \xi[k]} \text{CRLB}(\bm \xi[k]) \frac{\partial \vect\{\bH^*[k]\}}{\partial \bm \xi[k]}{\bigg{\}}}\nonumber \\
&=& \trace{\bigg{\{}}(\overline{\bA_\text{T}} \circ \bA_\text{R})\text{CRLB}(\bm \xi[k])(\overline{\bA_\text{T}} \circ \bA_\text{R})^*{\bigg{\}}}.
\end{eqnarray}
An important feature of this estimator is that the difference in performance given by the LS and the WLS estimators is more accentuated as the number of RF chains grows (if and only if the hybrid combiner is not built from orthonormal vectors). 
Furthermore, this estimator works even though the noise variance is unknown. This is because the noise covariance matrix $\bC$ can be written as $\bC = \sigma^2 \bC_\text{w}$, where $\bC_\text{w}$ models the coupling between the different combining vectors. Since the combining vectors are known at the receiver, this matrix $\bC_\text{w}$ is known as well. Therefore, we can write our estimator for the vector containing the channel gains, $\hat{\tilde{\bm \xi}}[k] \in \mathbb{C}^{\hat{L} \times 1}$ as
\begin{equation}
	\hat{\tilde{\bm \xi}}[k] = (\mathbf{\Upsilon}_{\{:,\hat{\cal T}\}}^*\bC_\text{w}^{-1}\mathbf{\Upsilon}_{\{:,\hat{\cal T}\}})^{-1}\mathbf{\Upsilon}_{\{:,\hat{\cal T}\}}^*\bC^{-1}_\text{w}\bsfy[k].
\end{equation}

When the noise variance is unknown, it follows from the Slepian-Bangs formula that the CRLB for the vector $\bm \xi[k]$ does not change, since the noise mean does not depend on the variance $\sigma^2$. Thus, the elements in the FIM modeling the coupling between the noise variance and the channel gains are zero. Therefore, the estimator of $\bm \xi[k]$ is efficient. In fact, it is also the ML estimator for this complex parameter vector, since it can be seen to maximize the Log-Likelihood Function (LLF) of the received signal. 
Nonetheless, there is no efficient estimator for both $\bm \xi[k]$ and $\sigma^2$, since the MVUE of $\sigma^2$ depends on the true value for $\bm \xi[k]$. 
Despite this, the MLE for $\sigma^2$ can still be found by setting the partial derivative of the LLF to zero. If we use the identities
\begin{equation}
	\frac{\partial \ln{\det\{\bC\}}}{\partial \sigma^2} = \trace{\bigg{\{}}\bC^{-1}\frac{\partial \bC}{\partial \sigma^2}{\bigg{\}}}
\end{equation}
and	
\be
\begin{split}
	\frac{\partial (\bsfy[k]-\mathbf{\Upsilon}_{\{:,\hat{{\cal T}}\}}\hat{\tilde{\bm \xi}}[k])^* \bC^{-1} (\bsfy[k]-\mathbf{\Upsilon}_{\{:,\hat{{\cal T}}\}}\hat{\tilde{\bm \xi}}[k])}{\partial \sigma^2} = -(\bsfy[k]-\mathbf{\Upsilon}_{\{:,\hat{{\cal T}}\}} \hat{\tilde{\bm \xi}}[k])^* \bC^{-1} \times \\ \frac{\partial \bC}{\partial \sigma^2} \bC^{-1} \times (\bsfy[k]-\mathbf{\Upsilon}_{\{:,\hat{{\cal T}}\}} \hat{\tilde{\bm \xi}}[k]),
\end{split}
\ee
it is possible to find that
\begin{equation}
	\hat{\sigma^2} = \frac{1}{M L_r}(\bsfy[k]-\mathbf{\Upsilon}\hat{\tilde{\bm \xi}}[k])^* \bC_\text{w}^{-1}(\bsfy[k]-\mathbf{\Upsilon}\hat{\tilde{\bm \xi}}[k]).
\end{equation}

Finally, we can consider the information coming from all the subcarriers to enhance the estimation of the noise variance. For the different subcarriers, we have the model
\begin{equation}
	\hat{\sigma^2}[k] = \sigma^2 + \nu[k],\qquad k = 0,\ldots,K-1,
\end{equation}
where $\nu[k]$ models the estimation error for the variance in the $k$-th subcarrier. We known that, in high SNR regime, the ML estimator is Gaussian distributed. Therefore, we have a set of $K$ linear models for the estimation of the variance. Since the model is linear on the parameter $\sigma^2$, we find that the ML of the variance is also the BLUE estimator, given by
\begin{equation}
	\hat{\sigma^2}_\text{ML} = \frac{1}{K}\mathbf{1}^T \bm \sigma,
\end{equation}
where $\bm \sigma \in \mathbb{R}^{K \times 1}$ is defined as $\bm \sigma = [\hat{\sigma^2}[0], \hat{\sigma^2}[1], \ldots, \hat{\sigma^2}[K-1]]^T$ and $\mathbf{1}$ is the $K\times 1$ column vector containing a $1$ in each position.

\begin{algorithm}
\caption{Simultaneous Weighted Orthogonal Matching Pursuit (SW-OMP)}\label{SWOMP}
\begin{algorithmic}[1]
\Procedure{SW-OMP($\bz[k]$,$\mathbf{\Phi}$,$\mathbf{\Psi}$,$\epsilon$)}{}
\State \textbf{Compute the equivalent observation matrix} \\ 
\qquad \qquad $\mathbf{\Upsilon} = \mathbf{\Phi} \mathbf{\Psi}$ 
\State \textbf{Initialize the residual vectors to the input signal vectors and support estimate}\\ 
\qquad \qquad $\bsfr[k] = \bsfy[k]$,\quad $k = 0,\ldots, K-1$, $\hat{\cal T} = \{\emptyset\}$ 
\While {\text{MSE} $> \epsilon$} \\
\vspace*{-7mm}
\qquad \State \textbf{Distributed Projection} \\ 
\qquad \qquad \qquad $\bc[k] = \mathbf{\Upsilon}^* \bD_\text{w}^{-*}\bsfr[k]$, \quad $k = 0,\ldots,K-1$ 
\qquad \State \textbf{Find the maximum projection along the different spaces} \\ 
\qquad \qquad \qquad $p^\star = \underset{p}{\arg\,\max}\,\sum_{k=0}^{K-1}{|\{\bc[k]\}_p|}$ 
\qquad \State \textbf{Update the current guess of the common support} \\ 
\qquad \qquad \qquad $\hat{\cal T} = \hat{\cal T} \cup p^\star$ \\ 
\vspace*{-7mm}
\qquad \State \textbf{Project the input signal onto the subspace given by the support using WLS} \\ 
\qquad \qquad \qquad $\bx_{\hat{\cal T}}[k] = (\mathbf{\Upsilon}_{\{:,\hat{\cal T}\}}^*\bC_\text{w}^{-1}\mathbf{\Upsilon}_{\{:,\hat{\cal T}\}})^{-1}\mathbf{\Upsilon}_{\{:,\hat{\cal T}\}}^*\bC_\text{w}^{-1}\bsfy[k]$, $k = 0,\ldots,K-1$ \\
\qquad \qquad \qquad $\overline{\hat{\cal T}} \cup \hat{\cal T} = \{1,2,\ldots,G_\text{t} G_\text{r}\}$ , $\overline{\hat{\cal T}} \cap \hat{\cal T} = \{\varnothing\}$ \\ 
\vspace*{-7mm}
\qquad \State \textbf{Update residual}\\ 
\qquad \qquad \qquad	$\bsfr[k] = \bsfy[k]-\mathbf{\Upsilon}_{\{:,\hat{\cal T}\}}\hat{\tilde{\bm \xi}}[k]$ , where $\hat{\tilde{\bm \xi}}[k] = \bx_{\hat{\cal T}}[k]$, \quad $k = 0,\ldots,K-1$ \\ 
\vspace*{-7mm}
\qquad \State \textbf{Compute the current MSE} \\ 
\qquad \qquad \qquad $\text{MSE} = \frac{1}{KML_\text{r}}\sum_{k=0}^{K-1}{\br^*[k] \bC_\text{w}^{-1} \br[k]}$ \\ 
\vspace*{-6mm}
\EndWhile
\EndProcedure
\end{algorithmic}
\end{algorithm}

The derived estimator for the channel gains $\hat{\bm \xi}[k]$ can be used in the S-OMP algorithm. In fact, the larger the number of subcarriers, the smaller the estimation variance the ML estimator can achieve. Thereby, if the number of averaging subcarriers $K$ is large enough, the lack of knowledge of the sparsity level is not so critical because of two reasons: 1) the computation of the support is more precise due to noise averaging during the projection process, and 2) if the support is estimated correctly, a particular estimate of $\sigma^2$ will be very close to the true noise variance, such that the chosen halting criterion is optimal from the Maximum Likelihood perspective. It should be clear that the higher the covariance between adjacent noise components, the larger the performance gap between the S-OMP and the SW-OMP algorithm will be, which actually depends on the ratio between $\Nr$ ($\Nt$) and $L_\text{r}$ ($L_\text{t}$).

The modification of the S-OMP algorithm to include the MVU estimator for the channel gains, as well as the whitening matrix to estimate the support is provided in Algorithm~\ref{SWOMP}. Notice that the proposed algorithm performs both noise whitening and channel estimation, such that interferences due to noise correlation are mitigated.

\subsection{Subcarrier Selection Simultaneous Weighted-OMP + Thresholding}
Despite the use of a single, subcarrier-independent measurement matrix $\bm \Upsilon$ to estimate the frequency-domain channels, the algorithm presented in the previous section exhibits high computational complexity. The SW-OMP algorithm considers the distributed projection coming from every subcarrier; however, a trade-off between performance and computational complexity can be achieved if a small number of subcarriers $K_\text{p}<<K$ is used, instead. The problem amounts as to how to choose those subcarriers, since no quality measure is available beforehand. The ideal situation would require knowledge of the Signal-to-Noise Ratio (SNR), which is unknown so far. Nonetheless, we have access to different frequency-domain received vectors $\bsfy[k]$, $k = 0,1,\ldots,K-1$. Therefore, the quality measure to be used could be the $l_2$-norm of the different vectors. Thereby, the $K_\text{p}$ subcarriers having largest $l_2$-norm can be used to derive an estimate of the support of the already defined sparse channel vectors $\bh^\text{v}[k]$, $k = 0,\ldots,K-1$. 

The main problem concerning Matching Pursuit (MP) algorithms comes from the lack of knowledge of the channel sparsity $L$. For this reason, there is usually an iteration at which $L$ paths have been detected but the estimate of the average residual energy is a little larger than the noise variance itself. This makes the algorithm find additional paths which are not actually contained in the MIMO channel. These paths usually have low power, and a pruning procedure is needed to filter out these undesired components. 

\vspace*{-1cm}
\begin{algorithm}
\caption{Subcarrier Selection Simultaneous Weighted Orthogonal Matching Pursuit + Thresholding (SS-SW-OMP+Th)}\label{SSSWOMP}
\begin{algorithmic}[1]
\Procedure{SS-SW-OMP($\bsfy[k]$,$\mathbf{\Phi}$,$\mathbf{\Psi}$,$K_\text{p}$,$\beta$,$\epsilon$)}{}
\State \textbf{Initialize counter, set of subcarriers and residual vectors} \\ 
\qquad $i = 0$,\quad ${\cal K} = \{\emptyset\}$,\quad $\bsfr[k] = \bsfy[k]$, $k = 0,\ldots,K-1$ 
\State \textbf{Find the $K_\text{p}$ strongest subcarriers}
\qquad \While {$i \leq K_\text{p}$} \\
\qquad \qquad ${\cal K} = {\cal K} \cup \underset{k\not\in {\cal K}}{\arg\,\max}\,\|\bsfy[k]\|_2^2$ \\
\qquad \qquad $i = i + 1$ 
\EndWhile
\State \textbf{Compute the equivalent observation matrix} \\ 
\qquad \qquad $\mathbf{\Upsilon} = \mathbf{\Phi} \mathbf{\Psi}$ 
\While {\text{MSE} $> \epsilon$}
\qquad \State \textbf{Distributed Projection} \\ 
\qquad \qquad \qquad $\bc[k] = \mathbf{\Upsilon}^* \bD_\text{w}^{-*}\bsfr[k]$, \quad $k \in {\cal K}$ 
\qquad \State \textbf{Find the maximum projection along the different spaces} \\ 
\qquad \qquad \qquad $p^\star = \underset{p}{\arg\,\max}\,\sum_{k\in {\cal K}}{|\{\bc[k]\}_p|}$ 
\qquad \State \textbf{Update the current guess of the common support} \\ 
\qquad \qquad \qquad $\hat{\cal T} = \hat{\cal T} \cup p^\star$  
\qquad \State \textbf{Project the input signal onto the subspace given by the support using WLS} \\ 
\qquad \qquad \qquad $\bx_{\hat{\cal T}}[k] = (\mathbf{\Upsilon}_{\{:,\hat{\cal T}\}}^*\bC_\text{w}^{-1}\mathbf{\Upsilon}_{\{:,\hat{\cal T}\}})^{-1}\mathbf{\Upsilon}_{\{:,\hat{\cal T}\}}^*\bC_\text{w}^{-1}\bsfy[k]$, $k = 0,\ldots,K-1$ \\
\qquad \qquad \qquad $\bx_{\overline{\hat{\cal T}}} = \b0$, \qquad \qquad \qquad where \quad $\overline{\hat{\cal T}} \cup \hat{\cal T} = \{1,2,\ldots,G_\text{t} G_\text{r}\}$ , $\overline{\hat{\cal T}} \cap \hat{\cal T} = \{\varnothing\}$  
\qquad \State \textbf{Update residual}\\ 
\qquad \qquad \qquad	$\bsfr[k] = \bsfy[k]-\mathbf{\Upsilon}_{\{:,\hat{\cal T}\}}\hat{\tilde{\bm \xi}}[k]$ , where $\hat{\tilde{\bm \xi}}[k] = \bx_{\hat{\cal T}}[k]$, \quad $k = 0,\ldots,K-1$ 
\qquad \State \textbf{Compute the current MSE} \\ 
\qquad \qquad \qquad $\text{MSE} = \frac{1}{KML_\text{r}}\sum_{k=0}^{K-1}{\bsfr^*[k] \bC_\text{w}^{-1} \bsfr[k]}$ 
\EndWhile
\State \textbf{Thresholding based on maximum average power}\\
\qquad \qquad $P^{\star} = \underset{\ell}{\max}\,\frac{1}{K}\sum_{k=0}^{K-1}{|\{\hat{\bm \xi}[k]\}_\ell|^2}$, \qquad $\hat{p}_{\text{av},i} = \frac{1}{K}\sum_{k=0}^{K-1}{|\{\hat{\bm \xi}[k]\}_i|^2},\quad i = 1,\ldots,\hat{L}$\\
\qquad \qquad  $\hat{\cal T}_\text{Th} = \bigcup{i}  \, / \, \hat{p}_{\text{av},i} \geq \beta P^{\star}$, $i \in \hat{\cal T}$\\
\qquad \qquad $\hat{\bm \xi}[k] = \{\hat{\bm \xi}[k]\}_{\hat{\cal T}_\text{Th}}$, $k = 0,\ldots,K-1$
\EndProcedure
\end{algorithmic}
\end{algorithm}
\newpage

A reasonable way to prune the undesired paths is to remove those components whose power falls below a given threshold, which can be related to the average power of the component in the estimated sparse vectors having maximum average power. Let us denote this power by $P^{\star}$. Then, the threshold can be defined as $\eta = \beta P^{\star}$, $\beta \in (0,1)$.
The value $P^{\star}$ is taken as $P^{\star} = \underset{\ell}{\max}\,\frac{1}{K}\sum_{k=0}^{K-1}{|\{\hat{\bm \xi}[k]\}_\ell|^2}$. 
To keep the common sparsity property we must ensure that the channel support after thresholding remains invariant. For this purpose, we define a signal $\hat{\bp}_\text{av} \in \mathbb{C}^{\hat{L} \times 1}$ whose $i$-th component is given by $\hat{p}_{\text{av},i} = \frac{1}{K}\sum_{k=0}^{K-1}{|\{\hat{\bm \xi}[k]\}_i|^2},\quad i = 1,\ldots,\hat{L}$, such that $\hat{\bp}_{\text{av}}$ measures the average power along the different subbands of each spatial component in the quantized angle grid. The final support after thresholding $\hat{\cal T}_\text{Th}$ is defined as $\hat{\cal T}_\text{Th} = \bigcup_{i=1}^{\hat{L}}{i} \, / \, \hat{p}_{\text{av},i} \geq \beta P^{\star}$.
Therefore, the components in $\hat{\bm \xi}[k]$ indexed by $\hat{\cal T}_\text{Th}$ are the final channel gains estimates for each subcarrier. The modification of the proposed SW-OMP algorithm to reduce computational complexity and implement this pruning procedure is provided in Algorithm~\ref{SSSWOMP}.

\section{Results}
\label{section:results}
This section includes the main empirical results obtained with the two proposed algorithms, SW-OMP and SS-SW-OMP + Thresholding, and comparisons with other the frequency-domain channel estimation algorithms including SSAMP \cite{GaoDaiWan:Channel-estimation-mmWave-massiveMIMO:16} and DGMP \cite{GaoHuDai:Channel-estimation-mmWave-massiveMIMO-FS:16}. To obtain these results, we perform Monte Carlo simulations averaged over many trials to evaluate the normalized mean squared error (NMSE) and the ergodic rate as a function of SNR and number of training frames $M$. We also provide calculations of the computational complexity for the proposed algorithms in Table \ref{tab:complexity} and prior work in Table \ref{tab:complexity2}.

The typical parameters for our system configuration are summarized as follows. Both the transmitter and the receiver are assumed to use Uniform Linear Arrays (ULAs) with half-wavelength separation. Such a ULA has steering vectors obeying the expressions $\{\ba_\text{T}(\theta_\ell)\}_n = \sqrt{\frac{1}{N_\text{t}}} e^{\jj n \pi\cos{(\theta_\ell)}}, \quad n = 0,\ldots,N_\text{t}-1$ and $\{\ba_\text{R}(\phi_\ell)\}_m = \sqrt{\frac{1}{N_\text{r}}} e^{\jj m \pi\cos{(\phi_\ell)}}, \quad m= 0,\ldots,N_\text{r}-1$. 
We take $N_\text{t}=N_\text{r}=32$ and $G_\text{t}=G_\text{r}=64$. The phase-shifters used in both the transmitter and the receiver are assumed to have $N_\text{Q}$ quantization bits, so that the entries of the training vectors $\fmtr$, $\wmtr$, $m = 1,2,\ldots,M$ are drawn from a set ${\cal A} = \left\{0,\frac{2\pi}{2^{N_\text{Q}}},\ldots,\frac{2\pi (2^{N_\text{Q}}-1)}{2^{N_\text{Q}}} \right\}$. The number of RF chains is set to $L_\text{t} = 1$ at the transmitter and $L_\text{r} = 4$ at the receiver. The number of OFDM subcarriers is set to $K = 16$.  

We generate channels according to \eqref{eqn:channel_model}  with the following parameters:
\begin{itemize}
\item The $L=4$ channel paths are assumed to be independent and identically distributed, with delay $\tau_\ell$ chosen uniformly from $[0,(N_\text{c}-1)T_\text{s}]$, with  $T_\text{s} = \frac{1}{1760}$ s, as in the IEEE 802.11ad wireless standard. 
\item The AoAs/AoDs are assumed to be uniformly distributed in $[0,\pi)$. 
\item The gains of each path are zero-mean complex Gaussian distributed such that $\mathbb{E}_k\{\|\bH[k]\|_F^2\} = \frac{N_\text{r}N_\text{t}}{\rho_\text{L}}$. 
\item The band-limiting filter $p_\text{rc}(t)$ is assumed to be a raised-cosine filter with roll-off factor of $0.8$. 
\item The number of delay taps of the channel is set to $N_\text{c} = 4$ symbols.    
\end{itemize}
The simulations we perform consider channel realizations in which the AoA/AoD are off-grid, i.e. do not correspond to the angles used to build the dictionary, and also the on-grid case, to analyze the loss due to the model mismatch.


\subsection{NMSE Comparison}
One performance metric  is the Normalized Mean Squared Error (NMSE) of a channel estimate $\hat{\bH}[k]$ for a given realization, defined as
		\begin{equation}
			\text{NMSE} = \frac{\sum_{k=0}^{K-1}{\|\hat{\bH}[k]-\bH[k]\|_F^2}}{\sum_{k=0}^{K-1}{\|\bH[k]\|_F^2}}.
		\end{equation}
The NMSE will be our baseline metric to compute the performance of the different estimators, and will be averaged over many channel realizations. 
The normalized CRLB (NCRLB) is also provided to compare the average performance of each algorithm with the lowest achievable NMSE, and will also be averaged over many channel realizations
		\begin{equation}
			\overline{\text{NCRLB}} = \frac{\sum_{k=0}^{K-1}{\overline{\text{CRLB}}(\vect\{\bH[k]\})}}{\sum_{k=0}^{K-1}{\|\bH[k]\|_F^2}}.
		\end{equation}

\begin{figure}[h!]
\centering
\begin{tabular}{cccc}
{\includegraphics[width=0.5\textwidth]{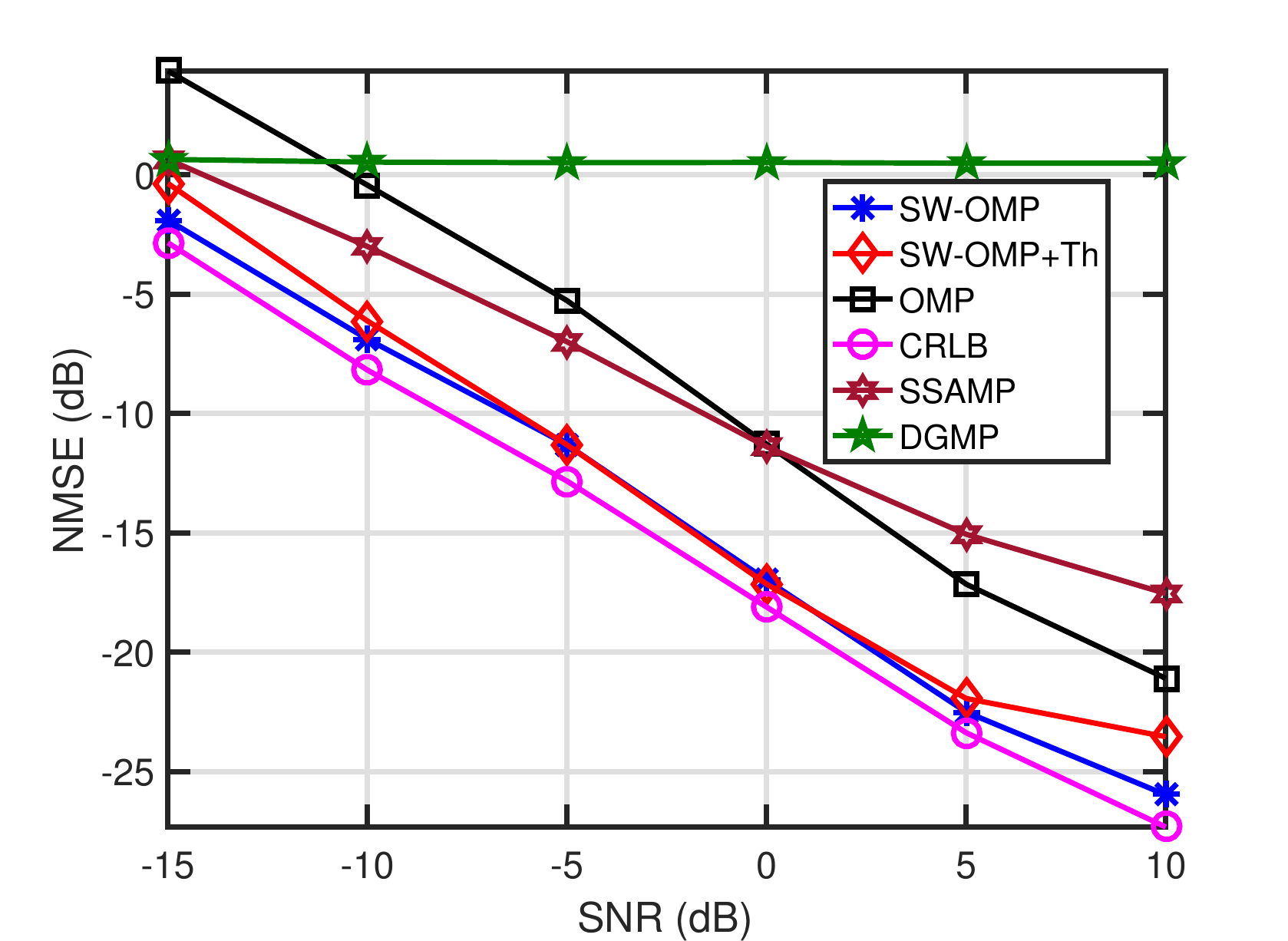}} & {\includegraphics[width=0.5\textwidth]{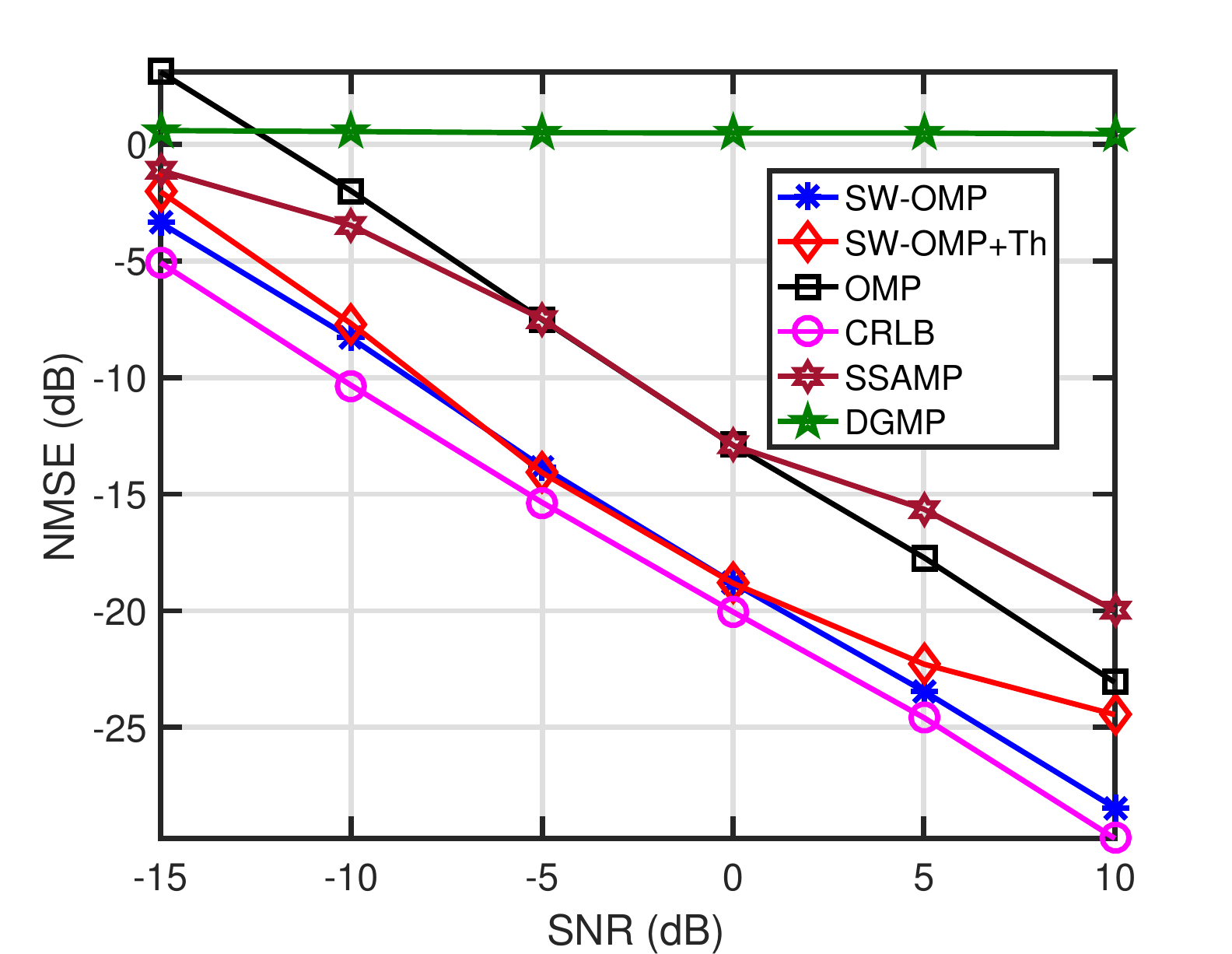}}\\
(a) & (b)
\end{tabular}
\caption{Evolution of the NMSE versus SNR for the different frequency-domain algorithms when the AoA/AoDs are assumed to lie on the dictionary grid. The number of training frames is set to $M = 80$ (a) and $M = 120$ (b).}     
\label{fig:NMSEvsSNR}        
\end{figure}

We compare the average NMSE versus SNR obtained for the different channel estimation algorithms in Fig.~\ref{fig:NMSEvsSNR} for a practical SNR range of $-15$ dB to $10$ dB, on-grid AoA/AoDs, and two different values for the number of training frames. SW-OMP performs the best, achieving NMSE values very close to the NCRLB. SS-SW-OMP+Th performs similarly to SW-OMP, although there is some performance loss due to the fact that SS-SW-OMP+Th does not employ every subcarrier to estimate the common support of the sparse channel vectors. In SS-SW-OMP+Th, the number of selected subcarriers for the estimation of the support was set to $K_\text{p} = 4$ for illustration and the parameter $\beta$ was chosen as $\beta = 0.025$, which is a reasonably small value to filter out undesired components in the sparse channel estimate. 
From the curves shown in Fig.~\ref{fig:NMSEvsSNR}, OMP performs poorly for all the SNR range due to the fact that it is not designed to process several vectors which are sparse in a common vector basis. Exploiting common sparsity provides an NMSE reduction of approximately $7$ dB, although there are slight variations depending on the SNR. This improvement comes at the cost of a higher offline computational complexity in the proposed algorithms in comparison with OMP, as we show in Section \ref{sec:complexity}. 

We also observe that the DGMP algorithm from \cite{GaoHuDai:Channel-estimation-mmWave-massiveMIMO-FS:16} performs the worst, due to the fact that it has been designed to estimate near-LoS mmWave MIMO channels. Since it only estimates a single path, the estimation error for NLOS channels is large. The algorithm SSAMP is also shown for comparison. We can see that, at low SNR regime, the information on the common support is enough to outperform the OMP algorithm, but not at the high SNR regime. This comes from the fact that the algorithm is estimating more than a single path per iteration. Provided that the dictionary matrices are not square in this setting, the redundancy between columns in the transmit and receive array matrices makes it difficult to properly estimate more than a support index per iteration.

\begin{figure}[h!]
\centering
\begin{tabular}{cccc}
{\includegraphics[width=0.45\textwidth]{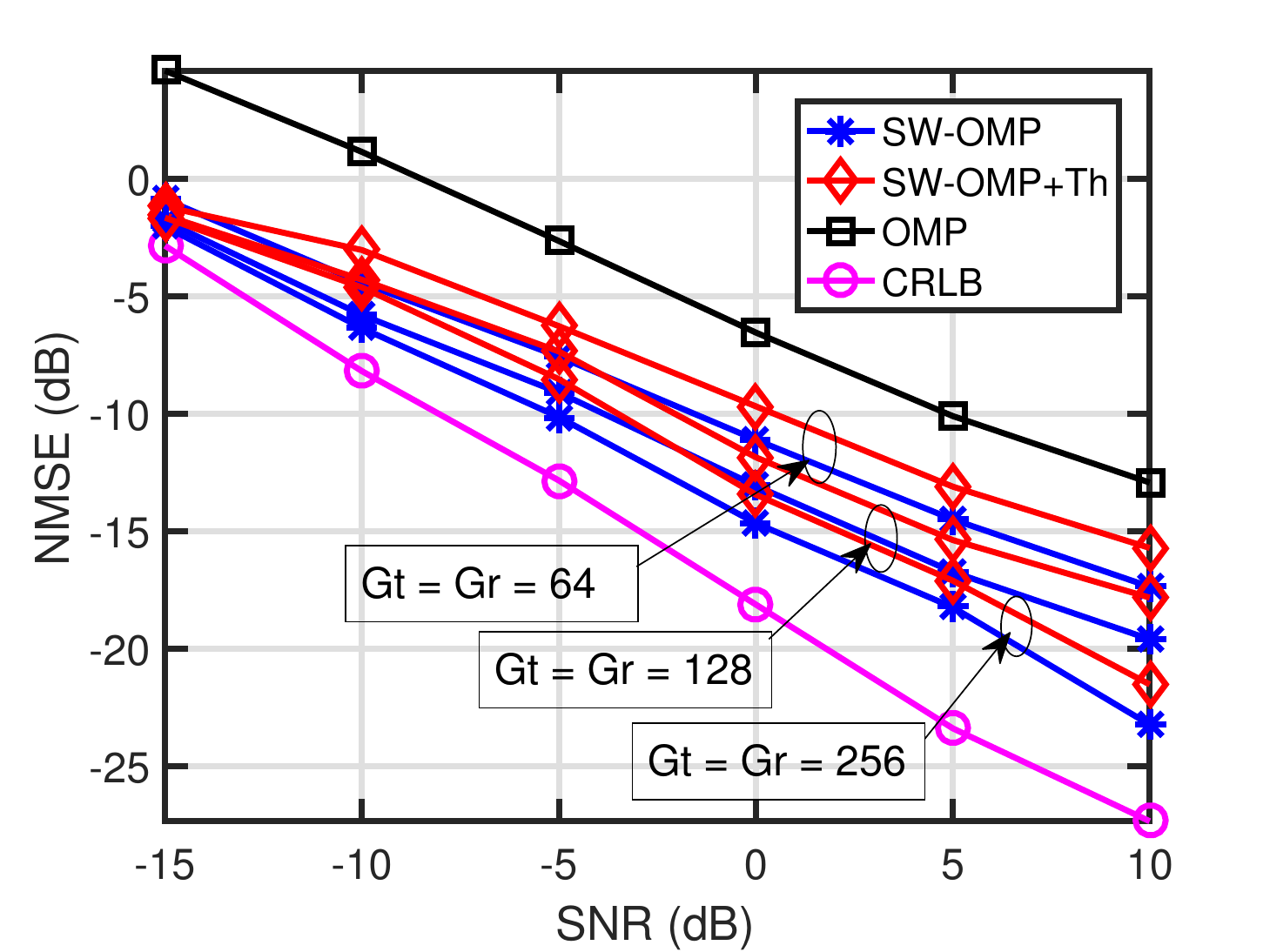}} & {\includegraphics[width=0.45\textwidth]{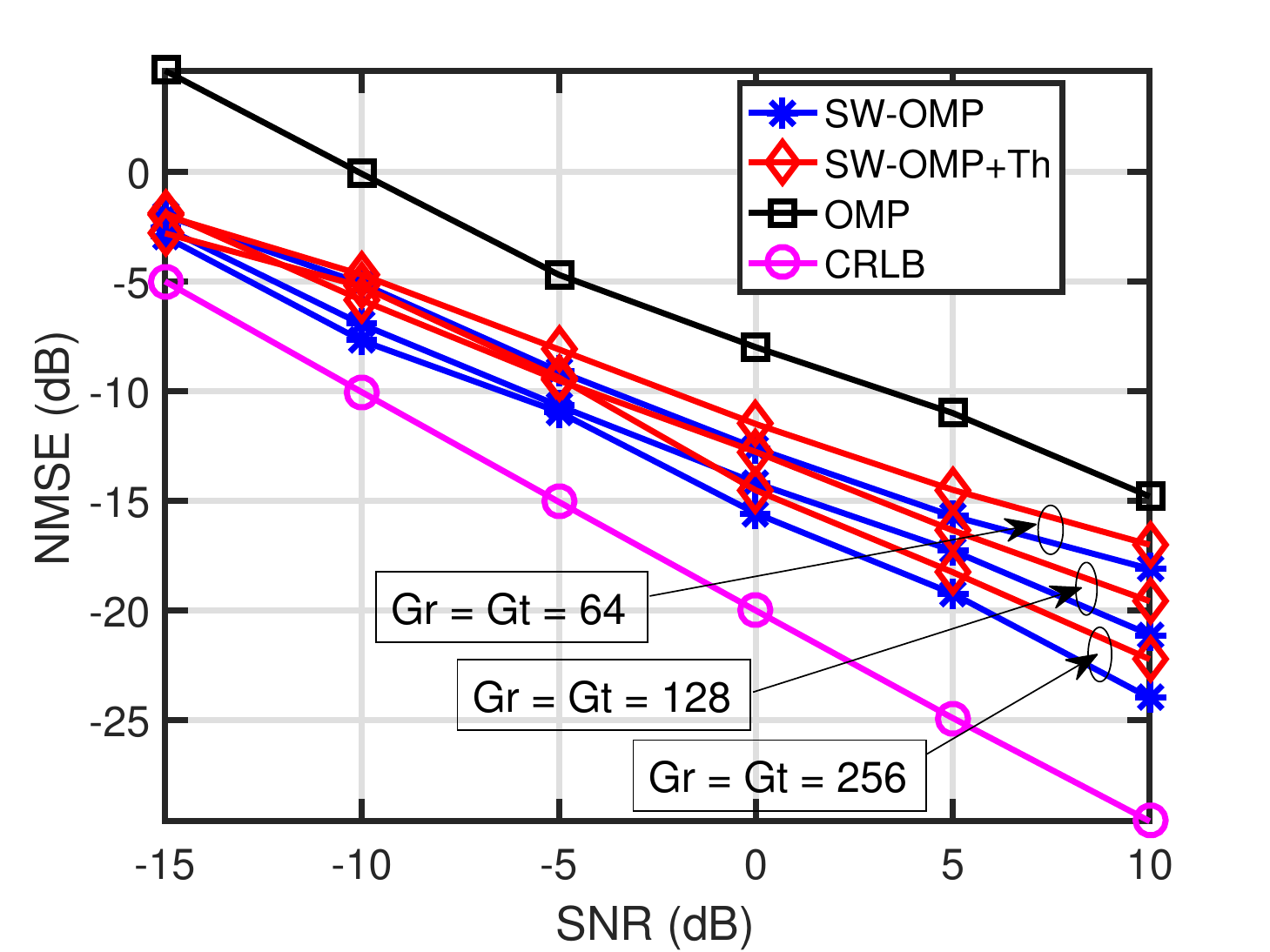}}\\
(a) & (b)
\end{tabular}
\caption{Evolution of the NMSE versus SNR for the different frequency-domain algorithms. The number of training frames is set to $M = 80$ (a) and $M = 120$ (b). The AoD/AoA are generated from a continuous uniform distribution.}     
\label{fig:NMSEvsSNR_OffGrid}        
\end{figure}

The previous simulations showed the performance of the algorithms when the channel fits the  on-grid model, but it is also important to analyze the performance in a practical scenario, when the AoA/AoD do not fall within the quantized spatial grid. Fig.~\ref{fig:NMSEvsSNR_OffGrid} shows the performance of the different algorithms for off-grid angular parameters. It can be observed that when using $G_\text{t}=G_\text{r}=64$ there is a considerable loss in performance due to the use of a fixed dictionary to estimate the AoA/AoD. However, the estimation error is below $-10$ dB for values of SNR in the order of $0$ and beyond. On the other hand, provided that the SNR expected in mmWave communication systems is in the order of $-20$ dB up to $0$ dB, the large gap between the proposed algorithms and the NCRLB needs to be reduced. Increasing the size of the dictionary is one of the possible solutions, as shown  
by the curves corresponding to $G_\text{t}=G_\text{r}=128$ and $G_\text{t}=G_\text{r}=256$.

\begin{figure}[h!]
\centering
\includegraphics[width=0.45\columnwidth]{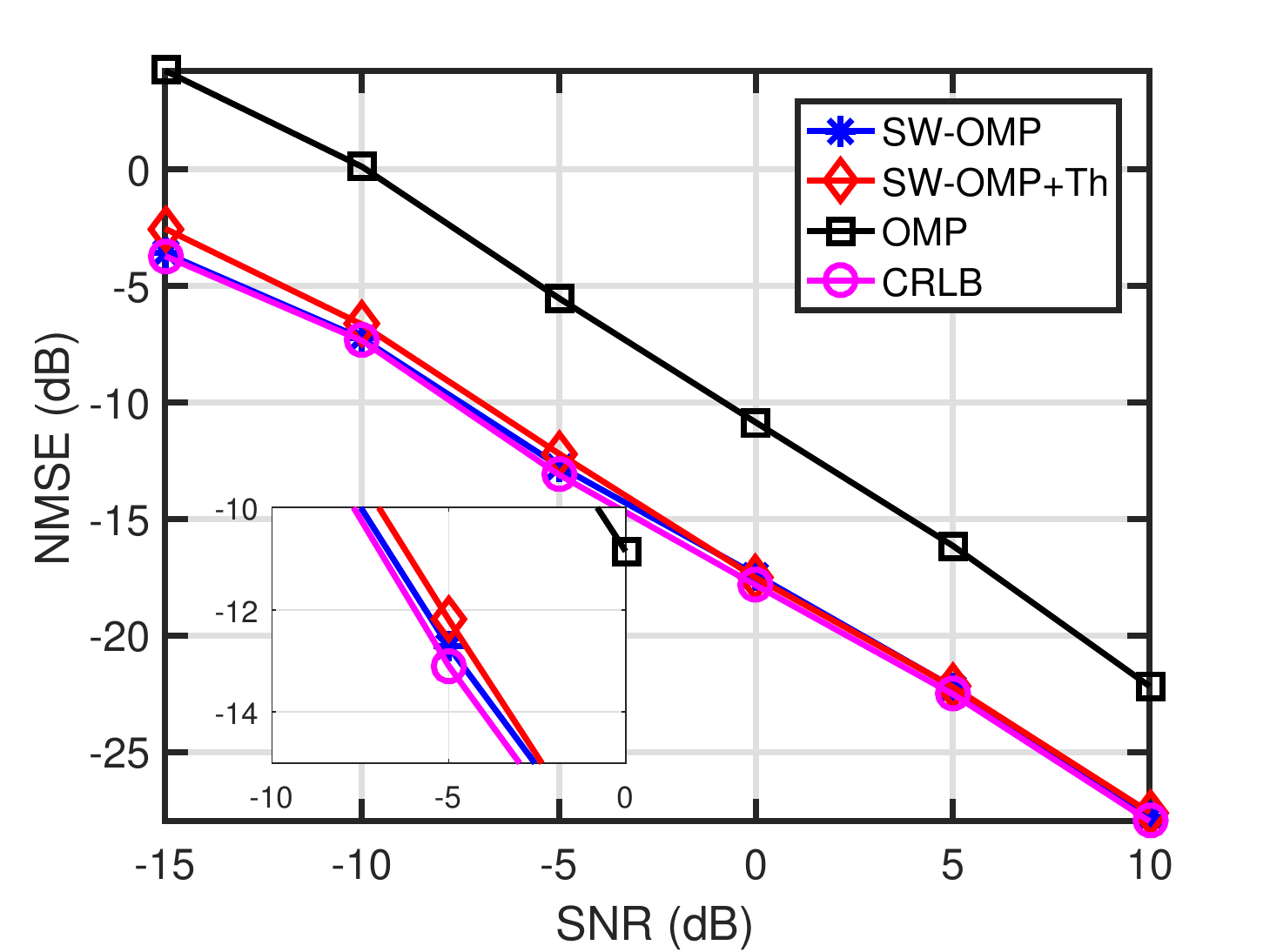} 
\caption{Comparison of evolution of the NMSE versus SNR for the different frequency-domain algorithms. The number of training frames is set to $M = 80$. The number of subcarriers is set to $K = 64$.}     
\label{fig:NMSEvsSNR2}        
\end{figure}

We show in Fig.~\ref{fig:NMSEvsSNR2} the performance of the different frequency-domain algorithms when increasing the number of subcarriers. The parameters for the simulation scenario are the same as in Fig.~\ref{fig:NMSEvsSNR}, however, the number of subcarriers is set to $K = 64$ in this case. $K_\text{p}$ is set to $32$ subcarriers and $\beta = 0.025\sigma^2$. Interestingly, both SW-OMP and SS-SW-OMP+Th are asymptotically efficient since they are both unbiased and attain the NCRLB. A magnified plot for $\text{SNR} = -5$ dB is also shown to clearly see the performance gap between the different algorithms and also the NCRLB. 

\begin{figure}[htb!]
    \centering
\begin{tabular}{cccc}
{\includegraphics[width=0.33\textwidth]{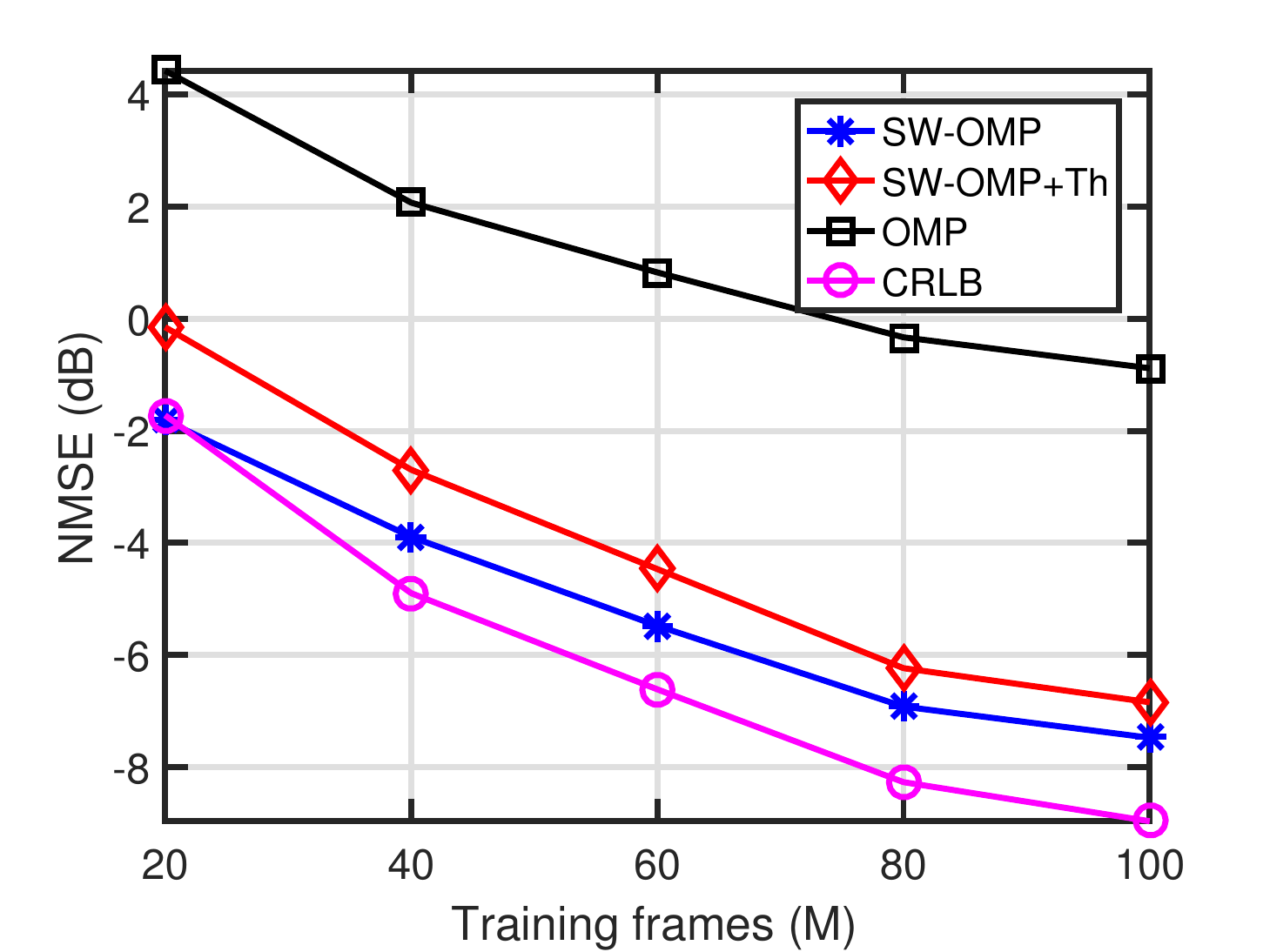}} & {\includegraphics[width=0.33\textwidth]{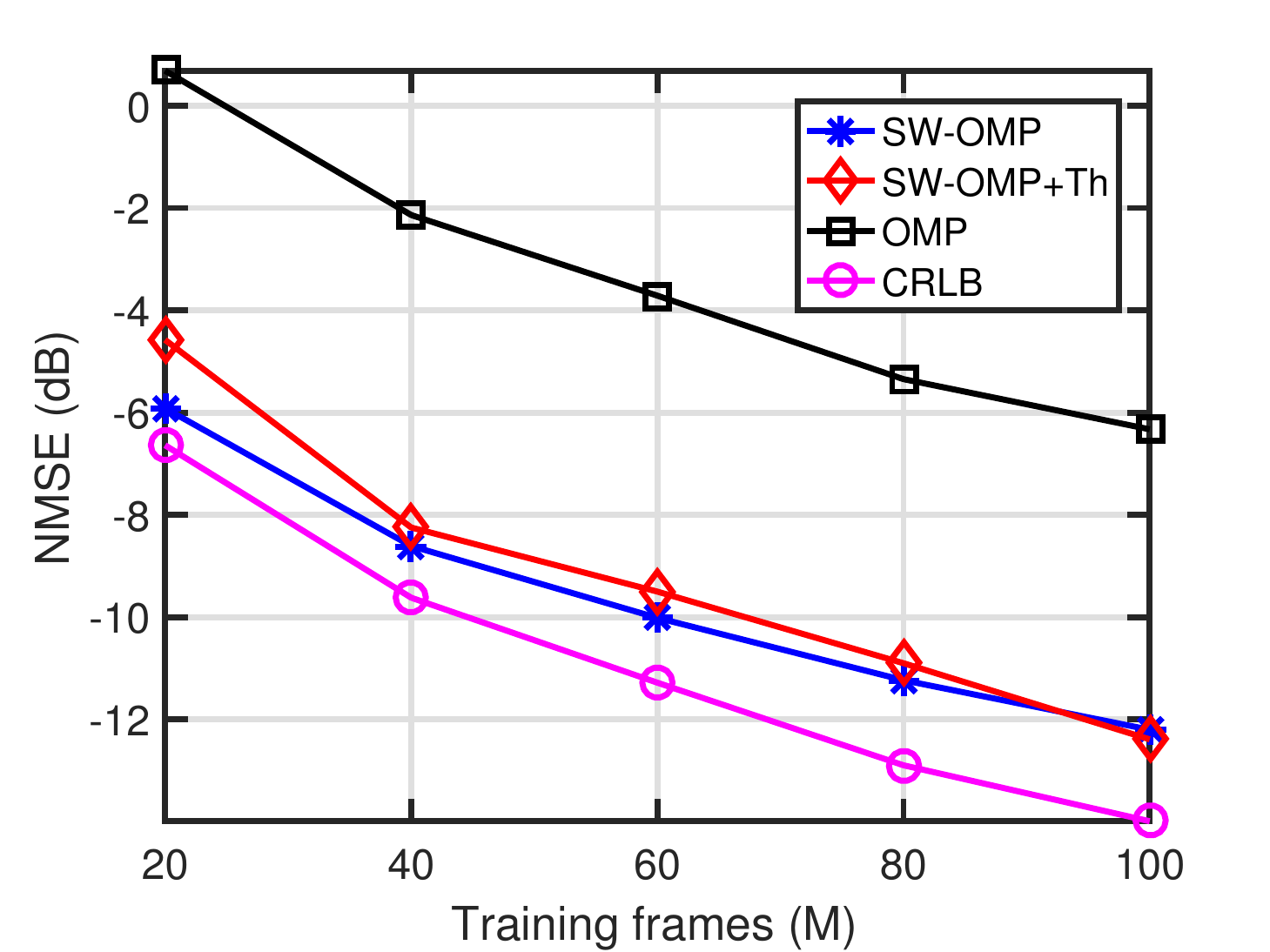}} & {\includegraphics[width=0.33\textwidth]{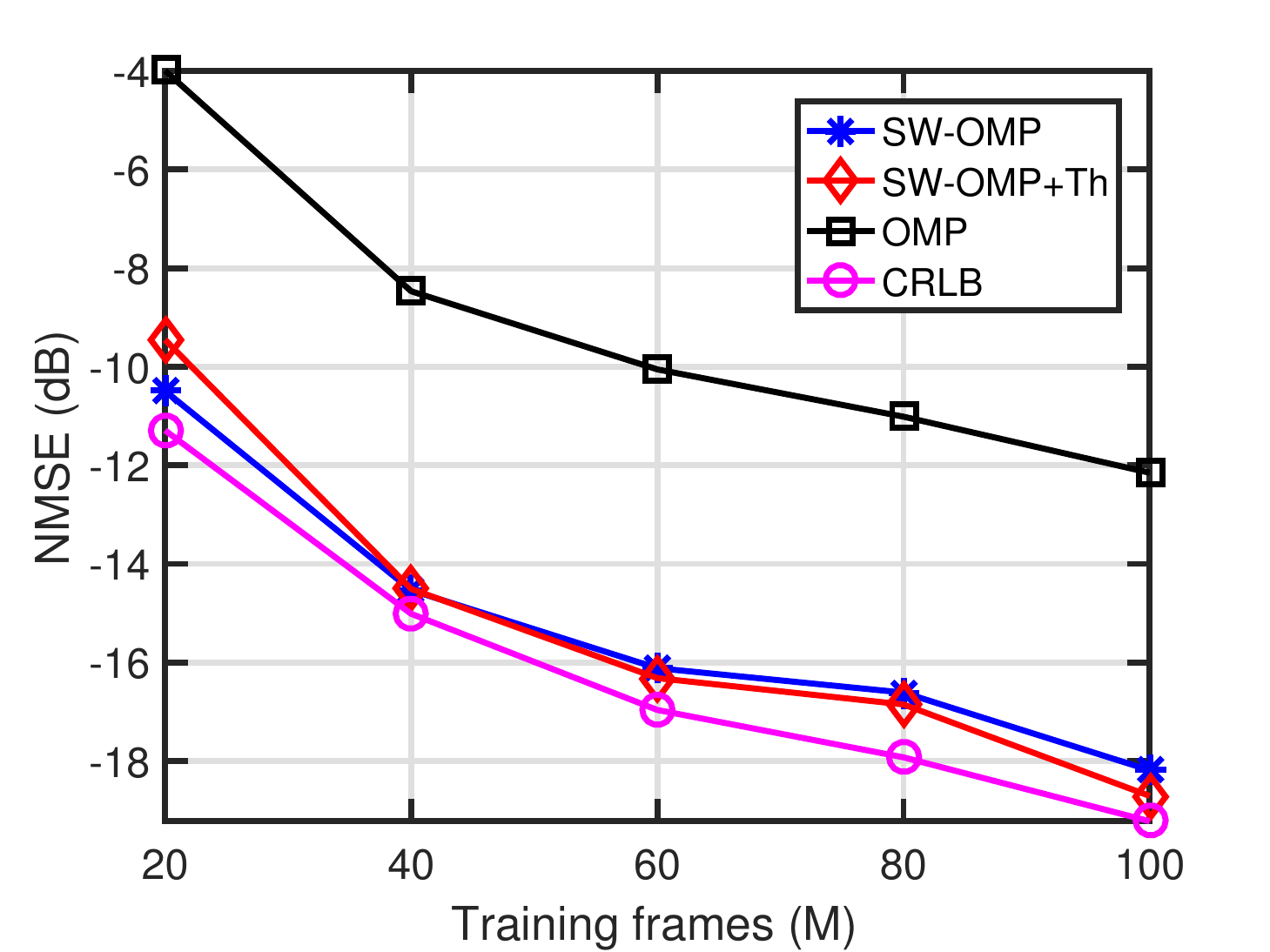}}\\
(a) & (b) & (c) \\
{\includegraphics[width=0.33\textwidth]{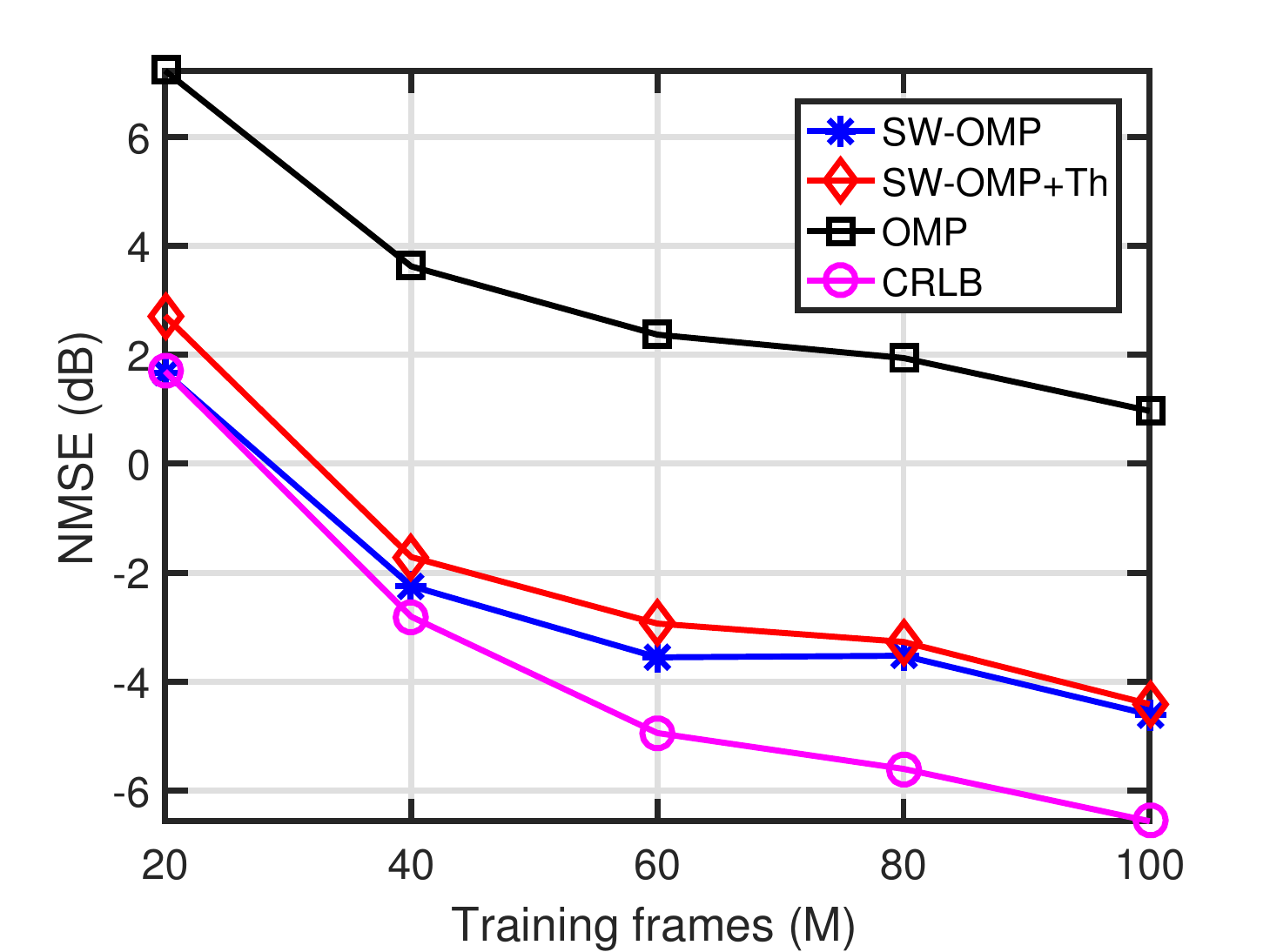}} & {\includegraphics[width=0.33\textwidth]{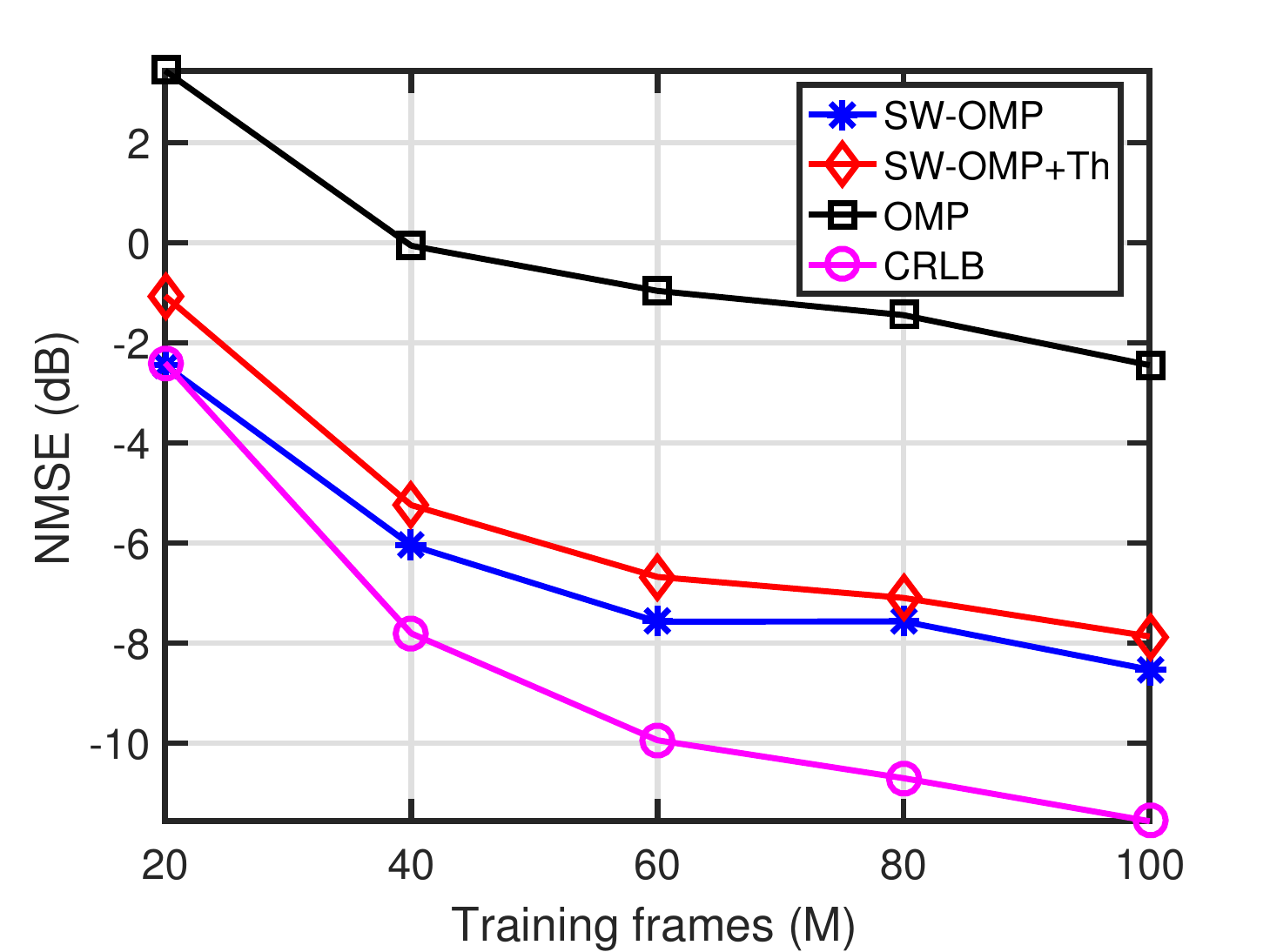}} & {\includegraphics[width=0.33\textwidth]{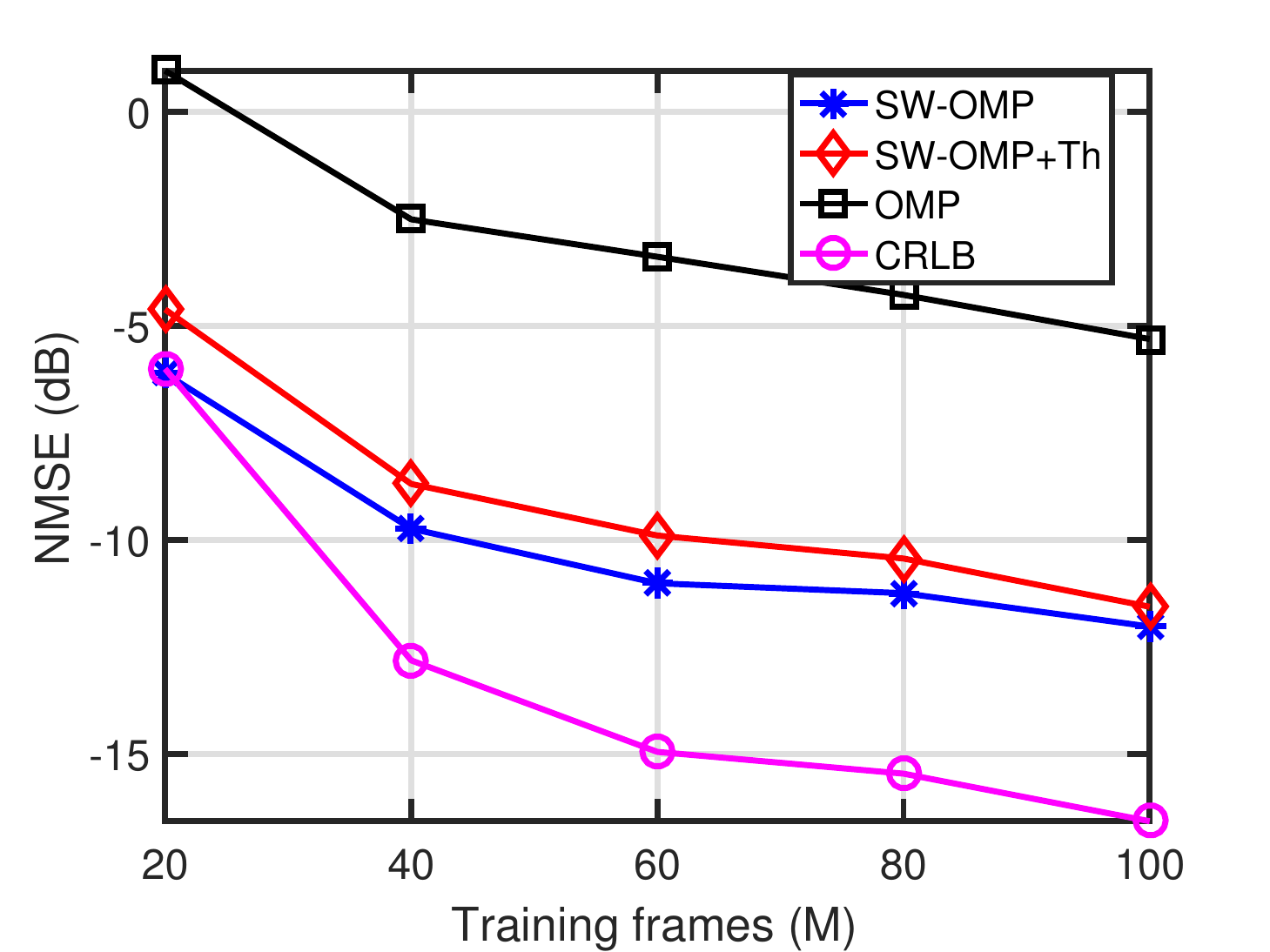}}\\
(d) & (e) & (f) \\
    \end{tabular}
    \caption{Comparison of evolution of the NMSE versus number of training frames $M$ at different SNR for the different channel estimation algorithms. The SNR is set to $-10$ dB (a) and (d), $-5$ dB (b) and (e) and $0$ dB (c) and (f). The curves in the first row consider on-grid angular parameters on the channel realizations, while the ones in the second row consider the off-grid case.}
    \label{fig:NMSEvsM}
\end{figure}

Fig.~\ref{fig:NMSEvsM} shows the average NMSE vs number of training frames $M$. The number of training frames $M$ is increased from $20$ to $100$. The remaining parameters in the simulation scenario are the same as in Fig.~\ref{fig:NMSEvsSNR}. Results are shown for channel realizations in which the angular parameters fall within the quantized angle grid and when they do not. The average performance of OMP algorithm is poor for all the considered cases, which comes from its inability to exploit the common support property shared by the different subchannels. SW-OMP can be observed to provide the best performance for all the values of $M$ and SNR. Furthermore, its performance lies closer to the NCRLB at higher SNR. The larger the number of training frames and the higher the SNR, the estimation of the support is more robust and gets closer to the actual one. The NCRLB provided assumes perfect sparse recovery, which is the reason why the estimation error does not lie within the bound. Nonetheless, in the on-grid case, if the number of training frames is large enough and the SNR is not low, the performance gap between SW-OMP and the NCRLB is lower than $1$ dB. The difference in performance between SS-SW-OMP+Th and SW-OMP reduces when either $M$ or SNR is increased.  As in Fig.~\ref{fig:NMSEvsSNR}, there is a big difference in performance between OMP and SW-OMP, depending on both the SNR and the number of frames. We also observe a high loss in performance when the angular parameters do not fall within the grid.

\subsection{Spectral efficiency comparison}
Another performance metric is the spectral efficiency,   computed assuming fully-digital precoding and combining using the $N_\text{s}$ dominant left and right singular vectors of the channel estimate. This gives $\Ns$ parallel channels each with gain $\lambda_n(\bH_\text{eff}[k])$, leading to average spectral efficiency
		\begin{equation}
			R = \frac{1}{K}\sum_{k=0}^{K-1} \sum_{n=1}^{N_\text{s}} \log\left( 1 + \frac{\rho}{K N_\text{s}\sigma^2} \lambda_n(\bH_\text{eff}[k])^2 \right).
		\end{equation}	
	
\begin{figure}[h!]
\centering
\begin{tabular}{cc}
\includegraphics[width=0.45\columnwidth]{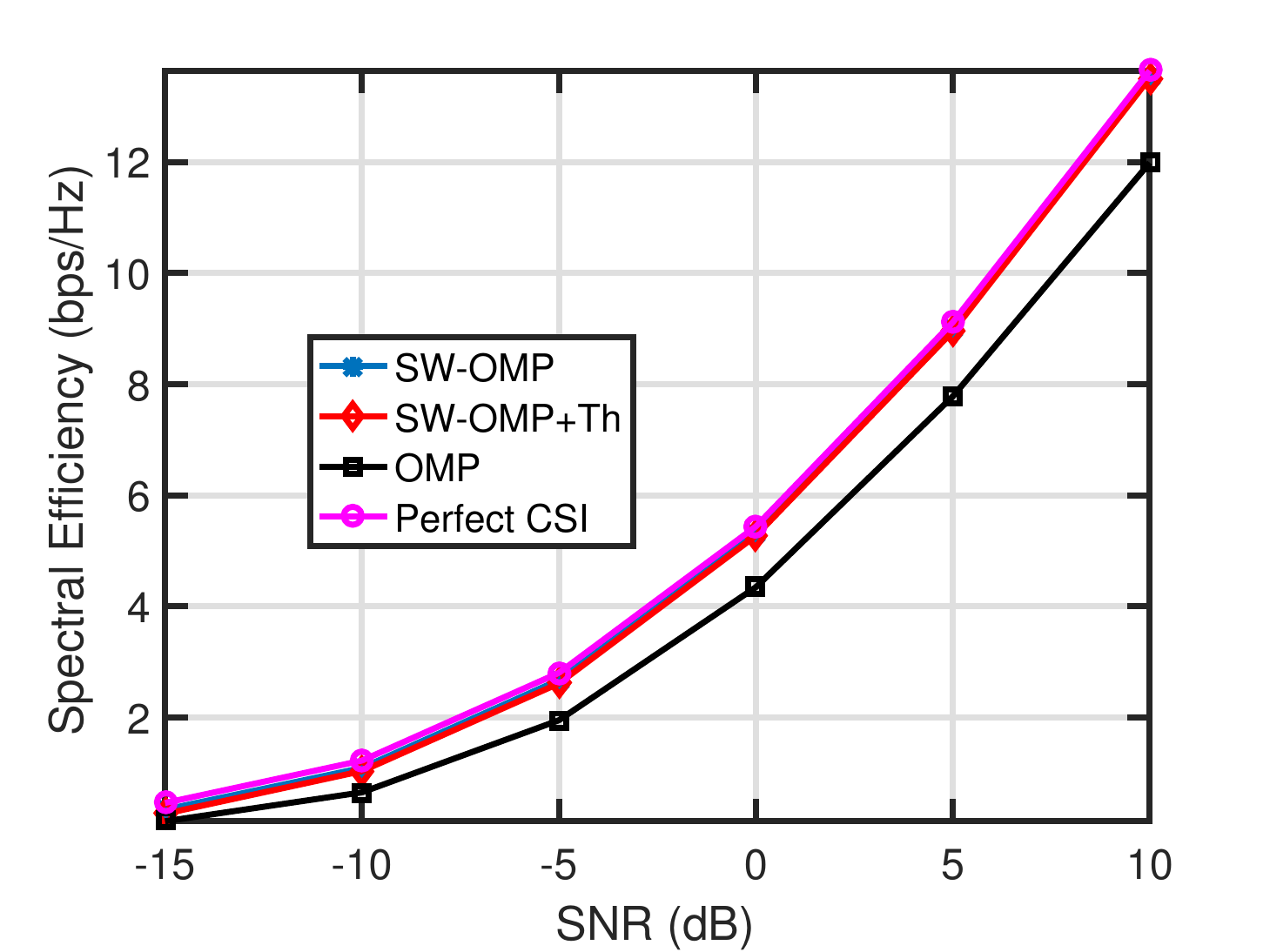} & \includegraphics[width=0.45\columnwidth]{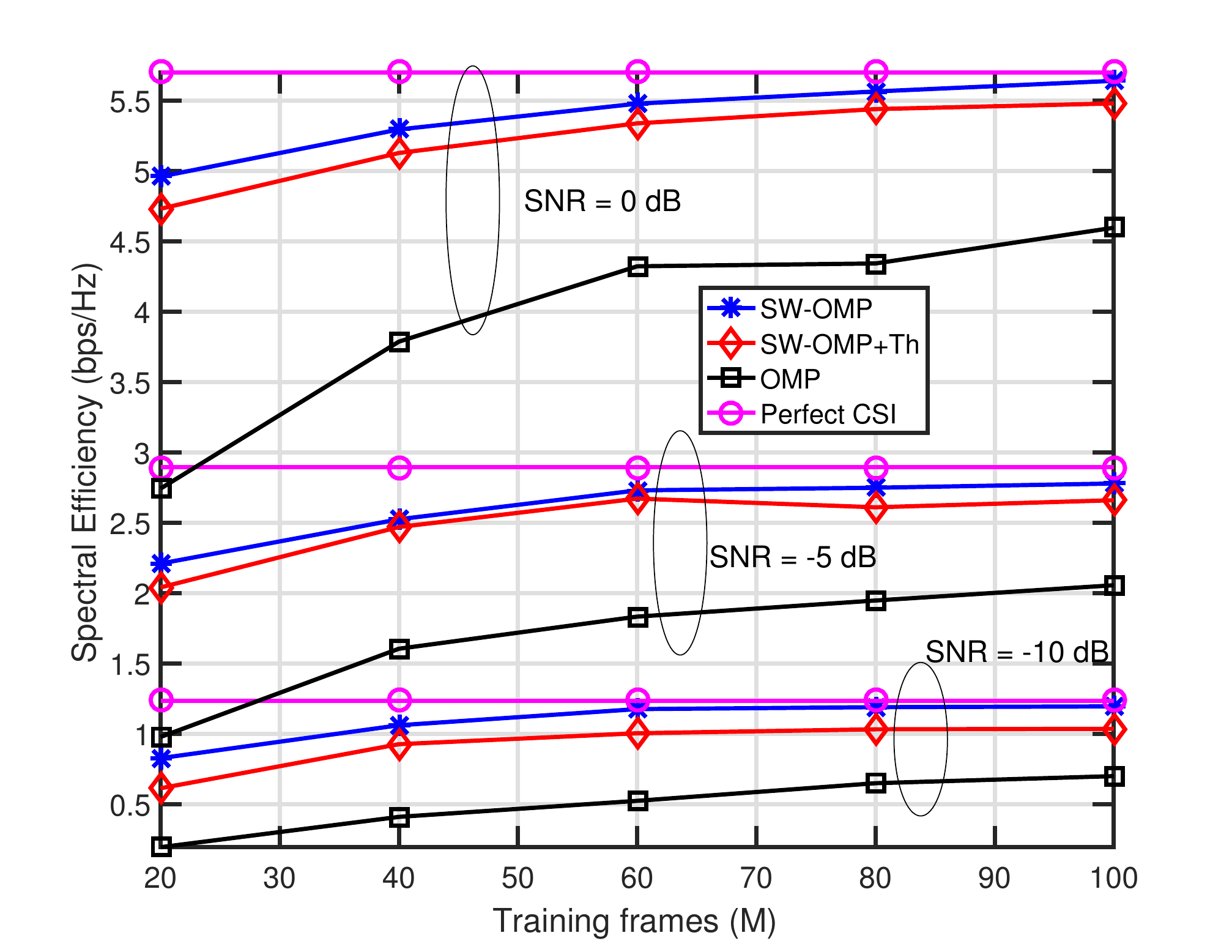} \\
(a) & (b) \\
\end{tabular}
\caption{(a) Evolution of the spectral efficiency versus SNR for the different frequency-domain algorithms. The number of training frames is set to $M = 80$. (b) Evolution of the spectral efficiency  versus number of training frames $M$ at different SNR for the different frequency-domain algorithms.}     
\label{fig:AchRate}        
\end{figure}

In Fig.~\ref{fig:AchRate}(a), we show the achievable spectral efficiency as a function of the SNR for the different channel estimation algorithms when off-grid AoA/AoDs are considered. The simulation parameters are the same as in Fig.~\ref{fig:NMSEvsSNR}. The difference in performance between OMP and either of our two proposed algorithms is approximately constant as the SNR increases, which comes from the fact that OMP does not force the channel estimates to share the same support. The two proposed algorithms perform very similarly for all the range of SNR, which is an indicator that $K_\text{p} < K$ subcarriers are enough to obtain a reliable channel estimate. Therefore, SS-SW-OMP+Th can be claimed to be a good trade-off between performance  and computational complexity.

Finally, Fig.~\ref{fig:AchRate}(b) shows the spectral efficiency as a function of the number of training frames $M$ for the different channel estimation algorithms. Comparisons are provided for $\text{SNR} = \{-10,-5,0\}$ dB. We observe that SW-OMP is the algorithm providing best performance, followed closely by SS-SW-OMP+Th, whilst OMP performs the worst. For low values of SNR, there is a noticeable gap between SW-OMP and the perfect CSI case, which becomes smaller as $M$ increases. For small values of $M$, the performance of the sparse recovery algorithms is very poor. Conversely, using $M \geq 60$ frames does not result in a significant improvement in performance. Simulations also show that near-optimal achievable rates can be achieved by using a reasonable number of frames, i.e., $60 \leq M \leq 100$.

\subsection{Computational complexity}\label{sec:complexity}	
The computational complexity for each step in the different algorithms is also provided in $j$ in Table~\ref{tab:complexity}. Since some steps can be performed before running the channel estimation algorithms, we will distinguish between on-line and off-line operations. Values are provided for a single iteration. In the case of the SSAMP algorithm in \cite{GaoDaiWan:Channel-estimation-mmWave-massiveMIMO:16} and the DGMP algorithm in \cite{GaoHuDai:Channel-estimation-mmWave-massiveMIMO-FS:16}, we take the notation used in the corresponding papers for frequency-domain vectors and measurement matrices.
	
\begin{table}[htbp]
  \centering
  \caption{Online Computational complexity of proposed algorithms}
    \begin{tabular}{ll}
    \toprule
    \multicolumn{2}{c}{SW-OMP} \\
    \midrule
    Operation & Complexity \\
    $(K) \times \bc[k] = \mathbf{\Upsilon}^*\bD_\text{w}^{-*} \br[k]$ & $\mathcal{O}(K (G_\text{r}G_\text{t}-(j-1)) L_\text{r}M)$ \\
    Maximum of $\sum_{k=0}^{N-1}{|\{\bc[k]\}_p|}$ & $\mathcal{O}(K (G_\text{r} G_\text{t}-(j-1)))$ \\
    $(K) \times \bx_{\hat{\cal T}}[k] = (\mathbf{\Upsilon}_{\{:,\hat{\cal T}\}}^*\bC_\text{w}^{-1}\mathbf{\Upsilon}_{\{:,\hat{\cal T}\}})^{-1}\mathbf{\Upsilon}_{\{:,\hat{\cal T}\}}^*\bC_\text{w}^{-1}\bz[k]$ & $\mathcal{O}(2 j^2 L_\text{r} M + j^3 + Kj L_\text{r} M)$ \\
    $(K) \times \br[k] = \bz[k] - \mathbf{\Upsilon}_{\{:,\hat{\cal T}\}}\hat{\bm \xi}[k]$  & $\mathcal{O}(K L_\text{r} M)$ \\
    $\text{MSE} = \frac{1}{K M L_\text{r}}\sum_{k=0}^{K-1}{\br^*[k] \bC_\text{w}^{-1}[k] \br[k]}$ & $\mathcal{O}(2 K L_\text{r} M)$ \\		
    Overall & $\mathcal{O}(L_\text{r} M (K(G_\text{r} G_\text{t}-(j-1)) +2j^2+Kj))$ \\
		\toprule
    \multicolumn{2}{c}{SS-SW-OMP+Thresholding} \\
		\midrule
		Operation & Complexity \\
		Find the $K_\text{p}$ strongest subcarriers & $\mathcal{O}(K L_\text{r} M)$ \\
    $(K_\text{p}) \times \bc[k] = \mathbf{\Upsilon}^*\bD_\text{w}^{-*} \br[k]$ & $\mathcal{O}(K_\text{p} (G_\text{r}G_\text{t}-(j-1)) L_\text{r}M)$ \\
    Maximum of $\sum_{k\in {\cal K}}{|\{\bc[k]\}_p|}$ & $\mathcal{O}(K_\text{p}( G_\text{r} G_\text{t}-(j-1)))$ \\
    $(K) \times \bx_{\hat{\cal T}}[k] = (\mathbf{\Upsilon}_{\{:,\hat{\cal T}\}}^*\bC_\text{w}^{-1}\mathbf{\Upsilon}_{\{:,\hat{\cal T}\}})^{-1}\mathbf{\Upsilon}_{\{:,\hat{\cal T}\}}^*\bC_\text{w}^{-1}\bz[k]$ & $\mathcal{O}(2 j^2 L_\text{r} M + j^3 + Kj L_\text{r} M)$ \\
    $(K) \times \br[k] = \bz[k] - \mathbf{\Upsilon}_{\{:,\hat{\cal T}\}}\hat{\bm \xi}[k]$  & $\mathcal{O}(K L_\text{r} M)$ \\
    $\text{MSE} = \frac{1}{K M L_\text{r}}\sum_{k=0}^{K-1}{\br^*[k] \bC_\text{w}^{-1}[k] \br[k]}$ & $\mathcal{O}(2 K L_\text{r} M)$ \\		
    Thresholding & $\mathcal{O}(K \hat{L})$ \\
    Overall & $\mathcal{O}(L_\text{r} M (K_\text{p} (G_\text{r} G_\text{t} -(j-1)) + 2j^2 + Kj))$ \\
    
    \bottomrule
    \end{tabular}%
  \label{tab:complexity}%
\end{table}%

\begin{table}[htbp]
  \centering
  \caption{Online Computational complexity of previously proposed algorithms}
    \begin{tabular}{ll}
		\toprule
    \multicolumn{2}{c}{OMP from \cite{VenAlkPreHeath:Time-domain-chan-estimation-wideband-hybrid:16}} \\
		\midrule
		Operation & Complexity \\
    $(K) \times \bc[k] = \mathbf{\Upsilon}^*[k] \br[k]$ & $\mathcal{O}(K (G_\text{r}G_\text{t} - (j-1)) L_\text{r}M)$ \\
    $(K) \times $ Maximum of $|\{\bc[k]\}_p|$ & $\mathcal{O}(K (G_\text{r} G_\text{t}-(j-1)))$ \\
    $(K) \times \bx_{\hat{\cal T}}[k] = \mathbf{\Upsilon}_{\{:,\hat{\cal T}\}}^\dag[k]\bz[k]$ & $\mathcal{O}(K(2 j^2 L_\text{r} M + j^3))$ \\
    $(K) \times \br[k] = \bz[k] - \mathbf{\Upsilon}_{\{:,\hat{\cal T}\}}[k]\hat{\bm \xi}[k]$  & $\mathcal{O}(K L_\text{r} M)$ \\
    $(K) \times \text{MSE} = \frac{1}{M L_\text{r}}\br^*[k] \br[k]$ & $\mathcal{O}(K L_\text{r} M)$ \\		
    Overall & $\mathcal{O}(K L_\text{r} M (G_\text{r} G_\text{t}-(j-1)) + 2j^2)$ \\
    


\toprule
    \multicolumn{2}{c}{SSAMP from \cite{GaoDaiWan:Channel-estimation-mmWave-massiveMIMO:16}} \\
    \midrule
    Operation & Complexity \\
    $(K) \times \ba_p = \mathbf{\Phi}_p^* \bb_p^{i-1}$ & $\mathcal{O}(K G_\text{r}G_\text{t} L_\text{r}M)$ \\
    Maximum of $\sum_{p=0}^{N-1}{\|\ba_p\|_2^2}$ & $\mathcal{O}(N G_\text{t} G_\text{r})$ \\
    $(K) \times \{\bt_p\}_{\Omega^{i-1} \cup \Gamma} = ((\mathbf{\Phi}_p)_{\Omega^{i-1} \cup \Gamma}^\dag \br_p$ & $\mathcal{O}(K (4 j^2 M L_\text{r} + (2 j)^3))$ \\
    Prune support $\Omega = \underset{\tilde{\Omega}}{\arg\,\max}\,\{\sum_{p=1}^{P}{\|\{\bt_p\}_{\tilde{\Omega}}\|_2^2}\}, |\tilde{\Omega}| = {\cal T}$ & $\mathcal{O}(2 K j)$ \\
    $(K) \times \{\bc_p\}_{\Omega} = (\{\mathbf{\Phi}_p\}_{\Omega})^\dag \br_p$ & $\mathcal{O}(K (2 j^2 M L_\text{r} + j^3))$ \\
    $(K) \times \bb_p = \br_p - \mathbf{\Phi}_p \bc_p$ & $\mathcal{O}(K L_\text{r} M G_\text{t} G_\text{r})$ \\
    Computation of total error $\sum_{p=0}^{K-1}{\|\bb_p\|_2^2}$ & $\mathcal{O}(K M L_\text{r})$ \\		
    Overall & $\mathcal{O}(2 K L_\text{r} M (G_\text{r} G_\text{t} + 6 j^2))$ \\


\toprule
    \multicolumn{2}{c}{DGMP from \cite{GaoHuDai:Channel-estimation-mmWave-massiveMIMO-FS:16}} \\
    \midrule
    Operation & Complexity \\
    $(K) \times \ba_p = \mathbf{\Upsilon}_p^* \br_p$ & $\mathcal{O}(K G_\text{r}G_\text{t} L_\text{r}M)$ \\
    Maximum of $\rho = \underset{\tilde\rho}{\arg\,\max}\,\sum_{p=0}^{N-1}{\|\ba_p\|_2^2}$ & $\mathcal{O}(K G_\text{t} G_\text{r})$ \\
    $(K) \times \{\alpha_p\}_{\rho} = (\{\mathbf{\Phi}_p\}_{\rho}^\dag \br_p$ & $\mathcal{O}(K (2 j^2 M L_\text{r} + j^3))$ \\	
    Overall & $\mathcal{O}(K L_\text{r} M (G_\text{r} G_\text{t} + 2 j^2))$ \\

    \bottomrule
    \end{tabular}%
  \label{tab:complexity2}%
\end{table}%

The computational complexity of SS-SW-OMP+Th is lower than in SW-OMP, since only a smaller number of projections are computed to estimate the channel support. It must be noticed that the complexity of SS-SW-OMP+Th is lower than that of OMP, since the matrix product $\bm \Upsilon^* \bC_\text{w}^{-1}$ can be computed before explicit channel estimation. The online computational complexity of SW-OMP is the same as that of OMP, since OMP performs the same projections as SW-OMP. Conversely, the offline computational complexities of both SW-OMP and SS-SW-OMP+Th are higher than those for the other algorithms, since the matrix product $\bm \Upsilon^*\bC_\text{w}^{-1}$ must be computed before explicit channel estimation. During estimation, only those rows corresponding to the estimate of the support must be selected to compute the channel gains. Nonetheless, OMP works in an isolated manner, independently solving each sparse recovery problem for each subcarrier, whereas SW-OMP exploits information that is common to every subcarrier. 

The complexity of the SSAMP algorithm proposed in \cite{GaoDaiWan:Channel-estimation-mmWave-massiveMIMO:16} is also shown for comparison. It is important to remark that this algorithm is based on the assumption that the frequency-selective hybrid precoders and combiners used during training are semiunitary. Therefore, this algorithm does not take into account the noise covariance matrix $\bC_\text{w}$, neither for the computation of the channel gains nor for the estimation of the support of the sparse channel matrices. 

We observe that the computational complexity of SW-OMP is lower than its SSAMP counterpart. This is because SSAMP exhibits an increase in complexity of at most $\mathcal{O}(4j^2)$ owing to the estimation of $j$ paths at the $j$-th iteration of SSAMP. This is because this algorithm uses an iteration index $i$ to estimate the sparsity level $L$, and a stage index $j$ to estimate the $j$ channel paths found at the current iteration. Afterwards, the support of the channel estimate is pruned to select the $j$ most likely channel paths. Therefore, at a given iteration $i$ and stage $j$, at most $2j = |\tilde{\Omega}^{i-1} \cup \Gamma|$ paths are estimated and then pruned, such that only $j$ paths are selected among the $2j$ candidates. The union of the sets $\Omega$ and $\Gamma$ comes from the possibility of finding new potential paths at the $i$-th iteration and the $j$-th stage. This is done by jointly considering the paths found at $(i-1)$-th iteration and the ones found in the $j$-th stage within the $i$-th iteration. Therefore, whereas both SW-OMP, SS-SW-OMP+Th and OMP estimate a single path at a given iteration $j$, SSAMP estimates at most $2j$ different paths by using LS. When computing the pseudoinverse during LS estimation, this results in an additional increase in complexity of $\mathcal{O}(4j^2)$, as shown in Table~\ref{tab:complexity2}. By contrast, as shown in Table~\ref{tab:complexity}, our proposed ML estimator for the channel gains exhibits computational complexity in the order of $\mathcal{O}(2j^2)$, thereby slightly reducing the number of operations. Further, both SW-OMP, SS-SW-OMP+Th and OMP do not compute the projections between every column of the measurement matrix and the received frequency-domain signals. Instead, a path that was already found in a previous iteration is not estimated again, thereby keeping computational complexity lower than that of SSAMP. 

Last, SSAMP considers the use of $K$ different measurement matrices for channel estimation. As discussed in Section~\ref{section:compressivechannelestimation}, this work considers analog-only training precoders and combiners to estimate the channel, which are constrained to be frequency flat. Consequently, the noise covariance matrix is constant across the different subcarriers. Besides, a frequency-flat data stream $\bs^{(m)} \in \mathbb{C}^{N_\text{s} \times 1}$ is transmitted at every subcarrier for a given training step $m$, $1\leq m \leq M$. Thereby, a single frequency-flat measurement matrix is used to jointly estimate the $K$ different subchannels. This results in a complexity reduction by a factor of $K$ at every iteration, both during the estimation of the support and during estimation of the channel gains. Thereby, the entire process of estimating the channel is greatly simplified.

While OMP does not require any off-line operation, both SW-OMP and SS-SW-OMP+Th need to compute the whitened measurement matrices $\bm \Upsilon_\text{w}= \bm \bD_\text{w}^{-1}\Upsilon$ for the different subcarriers. The offline computation of $\bD_\text{w}^{-1}$ has complexity of ${\cal O}(\frac{M}{3}L_\text{r}^3)$, since $\bC_\text{w}^{-1}$ is a block diagonal matrix containing $M$ hermitian matrices. Therefore, the Cholesky decomposition of $\bC_\text{w}^{-1}$ is nothing but the block diagonal matrix containing the Cholesky decompositions of the covariance matrices for the different training frames. The cost of such decomposition for a matrix $\bA \in \mathbb{C}^{k \times k}$ is ${\cal O}(\frac{k^3}{3})$. The overall cost is calculated taking this individual cost into account. It is important to remark that this cost comes from the use of frequency-flat precoders/combiners and training symbols. This entails a reduction in computational complexity with respect to the case in which frequency-selective baseband combiners were used during the channel estimation stage. Nonetheless, in our simulations, only analog precoders and combiners are used during the training stage, therefore reducing the computational complexity by a factor of $K$.

\section{Conclusions}
\label{section:conclusions}

In this paper, we proposed two compressive channel estimation approaches suitable for OFDM-based communication systems. These two strategies are based on jointly-sparse recovery to exploit information on the common basis that is shared for every subcarrier. Our compressive approaches enable MIMO operation in mmWave systems since the different subchannels are simultaneously estimated during the training phase. Further, if there is no grid quantization error and the estimation of the support is correct, we showed that our algorithms are asymptotically efficient since they asymptotically attain the CRLB. In simulations, we found that only a small number of subcarriers provide a high probability of correct support detection, thus our estimators approach the CRLB with reduced computational complexity. The approaches were also found to work well even with off-grid parameters, and to outperform competitive frequency-domain channel estimation approaches. For future work, it would be interesting to  analytically calculate the minimum required number of subcarriers to guarantee a high probability of correctly recovering the support of the sparse channel vectors. It would also be interesting to study the effects of other impairments including array miscalibration, beam squint, and synchronization errors. 



\begin{thebibliography}{1}

\bibitem{mmWaveBook} T. Rappapport \textit{et al}, \textit{Millimeter Wave Wireless Communications}. Pearson Education, Inc., 2014.

\bibitem{HeaPreRanRohSay:Overview-Sig-Proc-mmWaveMIMO:16} R.W. Heath, N. Gonz\'{a}lez-Prelcic, S. Rangan, W. Roh, and A. M. Sayeed, "An overview of signal processing techniques for millimeter wave MIMO systems", \textit{IEEE J. Sel. Areas Commun.}, vol. 10, no. 3, pp. 436-453, April 2016.

\bibitem{Ahmed_MIMOprecoding_combining:CM2014} A. Alkhateeb, J. Mo, N. Gonz\'{a}lez-Prelcic, and R. W. Heath Jr., "MIMO precoding and combining solutions for millimeter-wave systems", \textit{IEEE Commun. Mag.}, vol. 52, no. 12, pp. 122-131, Dec 2014

\bibitem{Malloy2012} M. L. Malloy and R. D. Nowak, "Near-optimal adaptive compressed sensing", in \textit{Proc. Asil. Conf. Signals, Syst. Comp. (ASILOMAR)}, Pacific Grove, CA, 2012, pp. 1935-1939.

\bibitem{Malloy2012a} -, "Near-optimal compressive binary search", \textit{arXiv preprint arXiv:1306.6239}, 2012.

\bibitem{Iwen2012} M. Iwen and A. Tewfik, "Adaptive strategies for target detection and localization in noisy environments", \textit{IEEE Trans. Signal Process.}, vol. 60, no. 5, pp. 2344-2353, 2012.

\bibitem{AlkAyaLeuHea:Channel-Estimation-and-Hybrid:14} A. Alkhateeb, O. E. Ayach, G. Leus, and R. W. Heath Jr. "Channel estimation and hybrid precoding for millimeter wave cellular systems", \textit{IEEE J. Sel. Topics Signal Process.}, vol. 8, no. 5, pp. 831-846, Oct. 2014.

\bibitem{Ramasamy2012a} D. Ramasamy, S. Venkateswaran, and U. Madhow, "Compressive adaptation of large steerable arrays", in \textit{Information Theory and Applications Workshop (ITA), 2012}, Feb. 2012, pp. 234-239.

\bibitem{Ramasamy2012b} -, "Compressive tracking with 1000-element arrays: A framework for multi-Gbps mm wave cellular downlinks", in \textit{Proc. Annual Allerton Conference on Communication, Control, and Computing}, Oct. 2012, pp. 690-697.

\bibitem{Berraki2014} D. E. Berraki, S. M. D. Armour, and A. R. Nix, "Application of compressive sensing in sparse spatial channel recovery for beamforming in mmWave outdoor systems", in \textit{Proc. IEEE Wireless Communications and Networking Conference (WCNC)}, Apr. 2014, pp. 887-892.

\bibitem{Lee2014} J. Lee, G. Gye-Tae, and Y. H. Lee, "Exploiting spatial sparsity for estimating channel sof hybrid MIMO systems in millimeter wave communications", in \text{Proc. IEEE Globecom}, 2014.

\bibitem{RiaRusPreAlkHea:Hybrid-MIMO-Architectures:16} R. M\'{e}ndez-Rial, C. Rusu, N. Gonz\'{a}lez-Prelcic, A. Alkhateeb, and R. W. Heath Jr., "Hybrid MIMO architectures for millimeter wave communications", in \textit{Proc. IEEE Globecom}, 2014.

\bibitem{VenAlkPreHeath:Time-domain-chan-estimation-wideband-hybrid:16} K. Venugopal, A. Alkhateeb, N. Gonz\'{a}lez-Prelcic, and R. W. Heath Jr., "Time-domain channel estimation for wideband millimeter wave systems with hybrid architecture", in \textit{submitted to Int. Conf. Acoust. Speech and Sig. Proc. (ICASSP)}, Sept. 2016, pp. 1-5.

\bibitem{VenAlkGon:Channel-Estimation-for-Hybrid:17} K. Venugopal, A. Alkhateeb, N. Gonz\'{a}lez-Prelcic, and R. W. Heath Jr., "Channel estimation for hybrid architecture based wideband millimeter wave systems", \textit{IEEE Journal on Selected Areas in Communications}, to appear, 2017.

\bibitem{RodGonVen:A-Frequency-Domain-Approach-to-Wideband:17} J. Rodr\'{i}guez-Fern\'{a}ndez, N. Gonz\'{a}lez-Prelcic, K. Venugopal, and R. W. Heath Jr., "A frequency-domain approach to wideband channel estimation in millimeter wave systems", in \textit{Proc. IEEE Int. Conf. Commun. (ICC)}, 2017.

\bibitem{GaoDaiWan:Channel-estimation-mmWave-massiveMIMO:16} Z. Gao, L. Dai, and Z. Wang, "Channel estimation for mmwave massive MIMO based access and backhaul in ultra-dense network", in \textit{Proc. IEEE Int. Conf. on Commun. (ICC)}, May 2016, pp. 1-6.

\bibitem{GaoHuDai:Channel-estimation-mmWave-massiveMIMO-FS:16} Z. Gao, C. Hu, L. Dai, and Z. Wang, "Channel estimation for millimeter-wave massive MIMO with hybrid precoding over frequency-selective fading channels", \textit{IEEE Communications Letters}, vol. 20, no 6, pp. 1259-1262, June 2016.

\bibitem{schniter_sparseway:2014} P. Schniter and A. Sayeed, "Channel estimation and precoder design for millimeter-wave communications: The sparse way", in \textit{Proc. Asilomar Conf. Signals, Syst., Comput.}, Nov. 2014, pp. 273-277.

\bibitem{AlkHea:Frequency-Selective-Hybrid:16} A. Alkhateeb and R. W. Heath Jr., "Freuqnecy selective hybrid precoding for limited feedback millimeter wave systems", \textit{IEEE Trans. Commun.}, vol. 64, no. 5, pp. 1801-1818, May 2016.

\bibitem{LarThoCud:Air-interface-design-and-ray-tracing:13} S. G. Larew, T. A. Thomas, M. Cudak, and A. Ghosh, "Air interface design and ray tracing study for 5g millimeter wave communicaitons", in \textit{Proc. IEEE Globecom Workshops (GC Wkshps)}, Dec. 2013.

\bibitem{Rap:TCOM:2015} T. S. Rappaport, G. R. MacCartney, M. K. Samimi, and S. Sun, "Wideband millimeter-wave propagation measurements and channel models for future wireless communication system design", \textit{IEEE Transactions on Communications}, vol. 63, no. 9, pp. 3029-3056, Sept. 2015.

\bibitem{TroGilStr:Simultaneous-sparse-approximation:05} J. A. Tropp, A. C. Gilbert, and M. J. Strauss, "Simultaneous sparse approximation via greedy pursuit", in \textit{Proc. IEEE Conf. Acous. Speech, and Signal Processing (ICASSP)}, 2005.

\bibitem{Kay:Fundamentals-of-Statistical-Signal:93} S. M. Kay, \textit{Fundamentals of Statistical Signal Processing, Volume I: Estimation Theory}. Prentice Hall PTR, 1993.

\end{thebibliography}

\end{document}